\def\ls{\lower4pt\hbox{${\buildrel < \over \sim}$}}
\def\gs{\lower4pt\hbox{${\buildrel > \over \sim}$}}

\documentclass[preprint2]{aastex}

\slugcomment{Submitted to {\it The Astrophysical Journal}}

\shorttitle{Multiwavelength Observations of 3C~66A in 2003 -- 2004}
\shortauthors{B\"ottcher et al.}

\begin{document}

\title{Coordinated Multiwavelength Observations of 3C~66A during the 
WEBT campaign of 2003 -- 2004\footnote{For questions regarding the 
availability of the data from the WEBT campaign presented in 
this paper, please contact the WEBT President Massimo Villata at 
{\tt villata@to.astro.it}}}

\author{M. B\"ottcher\altaffilmark{1}, J. Harvey\altaffilmark{1}, 
M. Joshi\altaffilmark{1}, M. Villata\altaffilmark{2}, C. M. Raiteri\altaffilmark{2}, 
D. Bramel\altaffilmark{3}, R. Mukherjee\altaffilmark{3}, T. Savolainen\altaffilmark{6},
W. Cui\altaffilmark{5}, G. Fossati\altaffilmark{4}, I. A. Smith\altaffilmark{4},
D. Able\altaffilmark{5},
H. D. Aller\altaffilmark{7}, M. F. Aller\altaffilmark{7}, 
A. A. Arkharov\altaffilmark{8}, 
K. Baliyan\altaffilmark{34}, D. Barnaby\altaffilmark{9}, 
A. Berdyugin\altaffilmark{6}, E. Ben\'\i tez\altaffilmark{14},
P. Boltwood\altaffilmark{10}, M. Carini\altaffilmark{9},
D. Carosati\altaffilmark{11}, S. Ciprini\altaffilmark{12}, 
J. M. Coloma\altaffilmark{13},
S. Crapanzano\altaffilmark{2}, J. A. de Diego\altaffilmark{14}, 
A. Di Paola\altaffilmark{40}, M. Dolci\altaffilmark{41},
J. Fan\altaffilmark{15},
A. Frasca\altaffilmark{16}, V. Hagen-Thorn\altaffilmark{17,18}, 
D. Horan\altaffilmark{37}
M. Ibrahimov\altaffilmark{19,20}, G. N. Kimeridze\altaffilmark{21}, 
Y. A. Kovalev\altaffilmark{36}, Y. Y. Kovalev\altaffilmark{36,42},
O. Kurtanidze\altaffilmark{21}, A. L\"ahteenm\"aki\altaffilmark{29}, 
L. Lanteri\altaffilmark{2}, V. M. Larionov\altaffilmark{17,18}, 
E. G. Larionova\altaffilmark{17}, E. Lindfors\altaffilmark{6}, 
E. Marilli\altaffilmark{16}
N. Mirabal\altaffilmark{22}, M. Nikolashvili\altaffilmark{21}, 
K. Nilsson\altaffilmark{6},
J. M. Ohlert\altaffilmark{23}, T. Ohnishi\altaffilmark{24}, 
A. Oksanen\altaffilmark{25},
L. Ostorero\altaffilmark{6,39}, G. Oye\altaffilmark{35},
I. Papadakis\altaffilmark{26,27}, M. Pasanen\altaffilmark{6},
C. Poteet\altaffilmark{9}, T. Pursimo\altaffilmark{35},
K. Sadakane\altaffilmark{24}, L. A. Sigua\altaffilmark{21},
L. Takalo\altaffilmark{6}, J. B. Tartar\altaffilmark{28}, 
H. Ter\"asranta\altaffilmark{29}, 
G. Tosti\altaffilmark{12}, R. Walters\altaffilmark{9}, K. Wiik\altaffilmark{6,38}, 
B. A. Wilking\altaffilmark{28}, W. Wills\altaffilmark{9}, 
E. Xilouris\altaffilmark{30}, 
A. B. Fletcher\altaffilmark{31}, M. Gu\altaffilmark{31,32,33}, 
C.-U. Lee\altaffilmark{31}, 
S. Pak\altaffilmark{31}, H.-S. Yim\altaffilmark{31}
}

\altaffiltext{1}{Astrophysical Institute, Department of Physics and Astronomy, \\
Clippinger 339, Ohio University, Athens, OH 45701, USA}
\altaffiltext{2}{Istituto Nazionale di Astrofisica (INAF), Osservatorio Astronomico di Torino,\\
Via Osservatorio 20, I-10025 Pino Torinese, Italy}
\altaffiltext{3}{Barnard College, Columbia University, New York, NY 10027, USA}
\altaffiltext{4}{Department of Physics and Astronomy, Rice University, MS 108, \\
6100 S. Main St., Houston, TX 77005-1892, USA}
\altaffiltext{5}{Department of Physics, Purdue University, 525 Northwestern Ave., \\
West Lafayette, IN 47907-1396, USA}
\altaffiltext{6}{Tuorla Observatory, University of Turku, 21500 Piikki\"o, Finland}
\altaffiltext{7}{Department of Astronomy, University of Michigan, 830 Dennison Building, \\
Ann Arbor, MI 48109-1090, USA}
\altaffiltext{8}{Pulkovo Observatory, Pulkovskoye Shosse, 65, 196140, St. Petersburg, Russia}
\altaffiltext{9}{Department of Physics and Astronomy, Western Kentucky University, \\
1 Big Red Way, Bowling Green, KY 42104, USA}
\altaffiltext{10}{Boltwood Observatory, 1655 Stittsville Main St., Stittsville, \\
Ontario, Canada, K2S 1N6}
\altaffiltext{11}{Osservatorio di Armenzano, Assisi, Italy}
\altaffiltext{12}{Osservatorio Astronomico, Universit\`a di Perugia, Via B.\ Bonfigli, \\
I-06126 Perugia, Italy}
\altaffiltext{13}{Agrupaci\`on Astronomica de Sabadell, Sabadell 08200, Spain}
\altaffiltext{14}{Instituto de Astronom\'{\i}a, UNAM, Apdo.\ Postal 70-264, \\
04510 M\'exico DF, Mexico}
\altaffiltext{15}{Center for Astrophysics, Guangzhou University, Guangzhou 510400, China}
\altaffiltext{16}{Osservatorio Astrofisico di Catania, Viale A.\ Doria 6, \\
I-95125 Catania, Italy}
\altaffiltext{17}{Astronomical Institute, St. Petersburg State University, \\
Universitetsky pr.\ 28, Petrodvoretz, 198504 St. Petersburg, Russia}
\altaffiltext{18}{Isaac Newton Institute of Chile, St. Petersburg Branch, \\
198504 St. Petersburg, Russia}
\altaffiltext{19}{Ulugh Beg Astronomical Institute, Academy of Sciences of Uzbekistan, \\
33 Astronomical Str., Tashkent 700052, Uzbekistan}
\altaffiltext{20}{Isaac Newton Institute of Chile, Uzbekistan Branch}
\altaffiltext{21}{Abastumani Observatory, 383762 Abastumani, Georgia}
\altaffiltext{22}{Department of Physics and Astronomy, Columbia University, \\
New York, NY 10027, USA}
\altaffiltext{23}{Michael Adrian Observatory, Astronomie-Stiftung Trebur, Fichtenstraße 7, \\
D-65468 Trebur, Germany}
\altaffiltext{24}{Astronomical Institute, Osaka Kyoiku University, Kashiwara-shi, \\
Osaka, 582-8582 Japan}
\altaffiltext{25}{Nyr\"ol\"a Observatory, Jyv\"askyl\"an Sirius ry, Kyllikinkatu 1, \\
40950 Jyv\"askyl\"a, Finland}
\altaffiltext{26}{Physics Department, University of Crete, 710 03 Heraklion, Crete, Greece}
\altaffiltext{27}{IESL, Foundation for Research and Technology-Hellas, \\
711 10 Heraklion, Crete, Greece}
\altaffiltext{28}{Department of Physics and Astronomy, University of Missouri-St.\ Louis, \\
8001 Natural Bridge Road, St.\ Louis, MO 63121, USA}
\altaffiltext{29}{Mets\"ahovi Radio Observatory, Helsinki University of Technology, \\
Mets\"ahovintie 114, 02540 Kylm\"al\"a, Finland}
\altaffiltext{30}{Institute of Astronomy and Astrophysics, NOA, I. Metaxa \& Vas. Pavlou str., \\
Palaia Penteli, Athens, Greece}
\altaffiltext{31}{Korea Astronomy \& Space Science Institute, 61-1 Whaam-Dong, Yuseong-Gu,\\
Daejeon 305-348, Korea}
\altaffiltext{32}{Shanghai Astronomical Observatory, Chinese Academy of Sciences, \\
80 Nandan Road, Shanghai 200030, China}
\altaffiltext{33}{National Astronomical Observatories, Chinese Academy of Sciences, \\
Beijing 100012, China}
\altaffiltext{34}{Physical Research Laboratory, Ahmedabad -- 3800 009, India}
\altaffiltext{35}{Nordic Optical Telescope, Apartado 474, E-38700 Santa Cruz de La Palma,\\
Santa Cruz de Tenerife, Spain}
\altaffiltext{36}{Astro Space Center, Profsoyuznaya St. 84/32, Moscow 117997, Russia}
\altaffiltext{37}{Harvard-Smithsonian Astrophysical Observatory, 60 Garden Street,\\
Cambridge, MA 02138, USA}
\altaffiltext{38}{Institute of Space and Astronautical Science, Japan Aerospace
Exploration Agency,\\
3-1-1 Yoshinodai, Sagamihara, Kanagawa 229-8510, Japan}
\altaffiltext{39}{Landessternwarte Heidelberg-K\"onigstuhl, K\"onigstuhl, \\
D-69117 Heidelberg, Germany}
\altaffiltext{40}{Istituto Nazionale di Astrofisica (INAF), Osservatorio Astronomico di Roma,\\
Via Frascati, Monteporzio Cantone, Roma, Italy}
\altaffiltext{41}{Istituto Nazionale di Astrofisica (INAF), Osservatorio Astronomico di Teramo,\\
Via Maggini, Teramo, Italy}
\altaffiltext{42}{National radio Astronomy Observatory, P. O. Box 2,\\
Green Bank, WV 24944, USA; Jansky Fellow}

\begin{abstract}
The BL~Lac object 3C~66A was the target of an extensive 
multiwavelength monitoring campaign from July 2003 through 
April 2004 (with a core campaign from Sept. -- Dec. 2003), 
involving observations throughout the electromagnetic 
spectrum. Radio, infrared, and optical observations were 
carried out by the WEBT-ENIGMA collaboration. At higher 
energies, 3C~66A was observed in X-rays (RXTE), and at 
very-high-energy (VHE) $\gamma$-rays (STACEE, VERITAS). In 
addition, the source has been observed with the VLBA at 9 epochs 
throughout the period September 2003 -- December 2004, including 
3 epochs contemporaneous with the core campaign.
 
A gradual brightening of the source over the course of the campaign 
was observed at all optical frequencies, culminating in a very 
bright maximum around Feb. 18, 2004. The WEBT campaign revealed 
microvariability with flux changes of $\sim 5$~\% on time scales 
as short as $\sim 2$~hr. The source was in a relatively 
bright state, with several bright flares on time scales of 
several days. The spectral energy distribution
(SED) indicates a $\nu F_{\nu}$ peak in the optical regime. A weak 
trend of optical spectral hysteresis with a trend of spectral softening 
throughout both the rising and decaying phases, has been found. On 
longer time scales, there appears to be a weak indication of a positive 
hardness-intensity correlation for low optical fluxes, which does not 
persist at higher flux levels. 

The 3 -- 10~keV X-ray flux of 3C~66A during the core campaign
was historically high and its spectrum very soft, indicating that 
the low-frequency component of the broadband SED extends beyond 
$\sim 10$~keV. No significant 
X-ray flux and/or spectral variability was detected. STACEE and 
Whipple observations provided upper flux limits at $> 150$~GeV 
and $> 390$~GeV, respectively.

The 22 and 43 GHz data from the 3 VLBA epochs made between September 2003 
and January 2004 indicate a rather smooth jet with only very moderate 
internal structure. Evidence for superluminal motion ($8.5 \pm 5.6 \, 
h^{-1}$~c) was found in only one out of 6 components, while the apparent 
velocities of all other components are consistent with 0. The radial 
radio brightness profile suggests a magnetic field decay $\propto r^{-1}$ 
and, thus, a predominantly perpendicular magnetic field orientation.
\end{abstract}


\section{Introduction}

BL Lac objects and flat-spectrum radio quasars (FSRQs) are 
active galactic nuclei (AGNs) commonly unified in the class 
of blazars. They exhibit some of the most violent high-energy
phenomena observed in AGNs to date. They are characterized
by non-thermal continuum spectra, a high degree of linear
polarization in the optical, rapid variability at all wavelengths, 
radio jets with individual components often exhibiting apparent 
superluminal motion, and --- at least episodically --- a significant 
portion of the bolometric flux emitted in $\gamma$-rays. 46 blazars 
have been detected and identified with high confidence in high energy 
($> 100$~MeV) gamma-rays by the EGRET instrument on board the 
{\it Compton Gamma-Ray Observatory} \citep{hartman99,mhr01}. 
The properties of BL~Lac objects and blazar-type FSRQs are 
generally very similar, except that BL~Lac objects usually 
show only weak emission or absorption lines (with equivalent 
width in the rest-frame of the host galaxy of $< 5$~\AA), if 
any. In 3C~66A (= 0219+428 = NRAO~102 = 4C~42.07), a weak Mg~II 
emission line has been detected by \cite{miller78}. This led 
to the determination of its redshift at $z = 0.444$, which was 
later confirmed by the detection of a weak Ly$\alpha$ line in 
the IUE spectrum of 3C~66A \citep{lanzetta93}. However, as recently 
pointed out by \cite{bramel05}, these redshift determinations are
actually still quite uncertain. In this paper, we do base our analysis
on a redshift value of $z = 0.444$, but remind the reader that some 
results of the physical interpretation should be considered as 
tentative pending a more solid redshift determination. 

In the framework of relativistic jet models, the low-frequency (radio
-- optical/UV) emission from blazars is interpreted as synchrotron
emission from nonthermal electrons in a relativistic jet. The
high-frequency (X-ray -- $\gamma$-ray) emission could either be
produced via Compton upscattering of low frequency radiation by the
same electrons responsible for the synchrotron emission \citep[leptonic
jet models; for a recent review see, e.g.,][]{boettcher02}, or 
due to hadronic processes initiated by relativistic protons 
co-accelerated with the electrons \citep[hadronic models, for 
a recent discussion see, e.g.,][]{muecke01,muecke03}. 

To date, 6 blazars have been detected at very high energies 
($> 300$~GeV) with ground-based air \v Cerenkov detectors 
\citep{punch92,quinn96,catanese98,chadwick99,aharonian02,horan02,holder03}.
All of these belong to the sub-class of high-frequency peaked BL~Lac 
objects (HBLs). The field of extragalactic GeV -- TeV astronomy is 
currently one of the most exciting research areas in astrophysics, as
the steadily improving flux sensitivities of the new generation of 
air \v Cerenkov telescope arrays and their decreasing energy thresholds 
\citep[for a recent review see, e.g.,][]{weekes02}, provides a growing 
potential to extend their extragalactic-source list towards intermediate 
and even low-frequency peaked BL~Lac objects (LBLs) with lower 
$\nu F_{\nu}$ peak frequencies in their broadband spectral energy 
distributions (SEDs). Detection of such objects at energies $\sim 40$ 
-- 100~GeV might provide an opportunity to probe the intrinsic 
high-energy cutoff of their SEDs since at those energies, $\gamma\gamma$ 
absorption due to the intergalactic infrared background is still 
expected to be small \citep[e.g.,][]{djs02}. 

3C~66A has been suggested as a promising candidate for 
detection by the new generation of atmospheric \v Cerenkov 
telescope facilities like STACEE or VERITAS \citep[e.g.][]{cg02}. 
\cite{neshpor98,neshpor00} have actually reported multiple detections 
of the source with the GT-48 telescope of the Crimean Astrophysical
Observatory, but those could not be confirmed by any other group 
so far \citep[see, e.g.][]{horan04}.

3C~66A is classified as a low-frequency peaked BL~Lac object (LBL),
a class also commonly referred to as radio selected BL~Lac objects. 
Its low-frequency spectral component typically peaks at IR -- UV
wavelengths, while the high-frequency component seems to peak
in the multi-MeV -- GeV energy range. Since its optical identification
by \cite{wills74}, 3C~66A has been the target of many radio, IR, 
optical, X-ray, and $\gamma$-ray observations in the past, although 
it is not as regularly monitored at radio frequencies as many other 
blazars due to problems with source confusion with the nearby radio 
galaxy 3C~66B (6'.5 from 3C~66A), in particular at lower (4.8 and 
8~GHz) frequencies \citep{aller94,takalo96}. 

The long-term variability of 3C~66A at near-infrared (J, H, and K bands) 
and optical (U, B, V, R, I) wavelengths has recently been compiled and
analyzed by \cite{fl99} and \cite{fl00}, respectively. Variability at
those wavelengths is typically characterized by variations over $\lesssim
1.5$~mag on time scales ranging from $\sim 1$~week to several years.
A positive correlation between the B -- R color (spectral hardness) 
and the R magnitude has been found by \cite{vagnetti03}. An 
intensive monitoring effort by \cite{takalo96} revealed evidence 
for rapid microvariability, including a decline $\sim 0.2$~mag 
within $\sim 6$~hr. Microvariability had previously been detected in
3C~66A by \cite{cm91} and \cite{diego97}, while other, similar efforts 
did not reveal such evidence \citep[e.g.][]{mg78,takalo92,xie92}. 
\cite{lainela99} also report on a 65-day periodicity of the source
in its optically bright state, which has so far not been confirmed 
in any other analysis.

3C~66A is generally observed as a point source, with no indication of
the host galaxy. The host galaxy of 3C~66A was marginally resolved by
\cite{wurtz96}. They found $R_{\rm Gunn} = 19.0^{\rm mag}$ for the host 
galaxy; the Hubble type could not be determined. 

In X-rays, the source has been previously detected by EXOSAT 
\citep{sambruna94}, EINSTEIN \citep{ww90}, ROSAT \citep{fossati98}, 
BeppoSAX \citep{perri03}, and {\it XMM-Newton} \citep{croston03}. 
It shows large-amplitude soft-X-ray variability among these 
various epochs of observation, with flux levels at 1~keV 
ranging from $\sim 0.4 \, \mu$Jy to $\sim 5 \, \mu$Jy and 
generally steep (energy index $\alpha > 1$) soft X-ray 
spectra below 1~keV. 3C~66A has also been detected in $> 100$~MeV
$\gamma$-rays by {\it EGRET} on several occasions, with flux 
levels up to $F_{> 100 {\rm MeV}} = (25.3 \pm 5.8) \times 
10^{-8}$~photons~cm$^{-2}$~s$^{-1}$ \citep{hartman99}. 

Superluminal motion of individual radio components of the jet
has been detected by \cite{jorstad01}. While the identification
of radio knots across different observing epochs is not unique,
\cite{jorstad01} favor an interpretation implying superluminal
motions of up to $\beta_{\rm app} \sim 19 \, h^{-1} \approx 
27$. This would imply a lower limit on the bulk Lorentz factor
of the radio emitting regions of $\Gamma \ge 27$. However, 
theoretical modeling of the non-simultaneous SED of 3C~66A 
\citep{ghisellini98} suggests a bulk Lorentz factor of the 
emitting region close to the core --- where the $\gamma$-ray 
emission is commonly believed to be produced --- of $\Gamma 
\sim 14$, more typical of the values obtained for other blazars 
as well.

In spite of the considerable amount of observational effort spent
on 3C~66A \citep[see, e.g.][for an intensive optical monitoring
campaign on this source]{takalo96}, its multiwavelength SED and 
correlated broadband spectral variability behaviour are still 
surprisingly poorly understood, given its possible VHE $\gamma$-ray 
source candidacy. The object has never been studied in a dedicated 
multiwavelength campaign during the life time of {\it EGRET}. There 
have been few attempts of coordinated multiwavelength observations. For 
example, \cite{worrall84} present quasi-simultaneous radio, IR, optical, 
and UV observations of 3C~66A in 1983, but observations at different 
wavelength bands were still separated by up to $\sim 2$~weeks, and 
no simultaneous higher-energy data were available. This is clearly 
inadequate to seriously constrain realistic, physical emission models, 
given the established large-amplitude variability on similar time 
scales and the importance of the high-energy emission in the broadband 
SED of the source.

For this reason, we have organized an intensive multiwavelength 
campaign to observe 3C~66A from July 2003 through April 
2004, focusing on a core campaign from Sept. -- Dec. 2003. 
In \S \ref{observations}, we describe the observations 
and data analysis and present light curves in the various 
frequency bands. Spectral variability patterns are 
discussed in \S \ref{variability}, and the results of 
our search for inter-band cross-correlations and time lags 
are presented in \S \ref{crosscorrelations}. Simultaneous 
broadband spectral energy distributions (SEDs) of 3C~66A 
at various optical brightness levels are 
presented in \S \ref{spectra}. In \S \ref{parameter_estimates} 
we use our results to derive estimates of physical parameters, 
independent of the details of any specific model.  We summarize 
in \S \ref{summary}. In a companion paper (Joshi \& B\"ottcher, 
2005, in preparation), we will use time-dependent leptonic 
models to fit the spectra and variability patterns found 
in this campaign, and make specific predictions concerning
potentially observable X-ray spectral variability patterns 
and $\gamma$-ray emission.

Throughout this paper, we refer to $\alpha$ as the energy 
spectral index, $F_{\nu}$~[Jy]~$\propto \nu^{-\alpha}$. A 
cosmology with $\Omega_m = 0.3$, $\Omega_{\Lambda} = 0.7$, 
and $H_0 = 70$~km~s$^{-1}$~Mpc$^{-1}$ is used. In this cosmology,
and using the redshift of $z = 0.444$, the luminosity distance 
of 3C~66A is $d_L = 2.46$~Gpc. 

\begin{figure}
\plotone{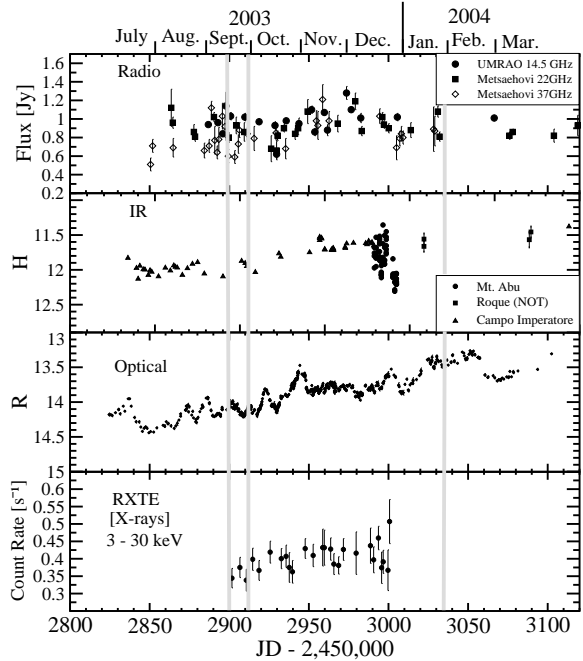}
\caption{Timeline of the broadband campaign on 3C~66A in 2003/2004.
The grey vertical lines indicate the epochs of the VLBA observations.}
\label{timeline}
\end{figure}

\section{\label{observations}Observations, data reduction, 
and light curves}

3C~66A was observed in a coordinated multiwavelength campaign 
at radio, near-IR, optical (by the WEBT-ENIGMA collaboration), 
X-ray, and VHE $\gamma$-ray energies during a core campaign 
period of September 2003 -- December 2003. The object is 
being continuously monitored at radio and optical wavelength
by several ongoing projects, and we present data in those wavelengths
regimes starting in July 2003. During the core campaign, we found
that the source was gradually brightening, indicating increasing
activity throughout the campaign period. For this reason, the
WEBT campaign was extended, and we continued to achieve good 
time coverage until early March 2004, and individual observatories
still contributed data through April 2004. The overall timeline 
of the campaign, along with the measured long-term light curves 
at radio, infrared, optical, and X-ray frequencies is 
illustrated in Fig. \ref{timeline}. Table \ref{observatories} 
lists all participating observatories which contributed data to 
this campaign. In this section, we will describe the individual 
observations in the various frequency ranges and outline the data 
reduction and analysis.

\begin{deluxetable}{lccc}
\tabletypesize{\scriptsize}
\tablecaption{List of observatories that contributed data to this campaign}
\tablewidth{0pt}
\tablehead{
\colhead{Observatory} & \colhead{Specifications} & \colhead{frequency / filters / energy range} &
\colhead{$N_{\rm obs}$}
}
\startdata
\multispan 4 \hss \bf Radio Observatories \hss \\
\noalign{\smallskip\hrule\smallskip}
UMRAO, Michigan, USA & 26 m  & 4.8, 8, 14.5 GHz & 46  \\
Mets\"ahovi, Finland & 14 m  & 22, 37 GHz       & 103 \\
RATAN-600, Russia    & 576m (ring) & 2.3, 3.9, 7.7, 11, 22 GHz & 17 \\
VLBA                 & 10x25 m & 2.3, 5, 8.4, 22, 43, 86 GHz & 9 \\
\noalign{\smallskip\hrule\smallskip}
\multispan 4 \hss \bf Infrared Observatories \hss \\
\noalign{\smallskip\hrule\smallskip}
Campo Imperatore, Italy     & 1.1 m & J, H, K  & 171 \\
Mt. Abu, India (MIRO)       & 1.2 m (NICMOS-3) & J, H, K' & 79 \\
Roque (NOT), Canary Islands & 2.56 m & J, H, K & 15 \\
\noalign{\smallskip\hrule\smallskip}
\multispan 4 \hss \bf Optical Observatories \hss \\
\noalign{\smallskip\hrule\smallskip}
Abastumani, Georgia (FSU)     & 70 cm         & R             & 210 \\
Armenzano, Italy              & 40 cm         & B, V, R, I    & 315 \\
Bell Obs., Kentucky, USA      & 60 cm         & R             & 12 \\
Boltwood, Canada              & 40 cm         & B, V, R, I    & 402 \\
Catania, Italy                & 91 cm         & U, B, V       & 835 \\
Crimean Astr. Obs., Ukraine   & 70 cm ST-7    & B, V, R, I    & 85 \\
Heidelberg, Germany           & 70 cm         & B, R, I       & 8 \\
Kitt Peak (MDM), Arizona, USA & 130 cm        & B, V, R, I    & 147 \\
Michael Adrian Obs., Germany  & 120 cm        & R             & 30 \\
Mt. Lemmon, Korea             & 100 cm        & B, V, R, I    & 399 \\
Mt. Maidanak, Uzbekistan      & 150 cm AZT-22 & B, R          & 1208 \\
Nyr\"ol\"a, Finland           & 40 cm SCT     & R             & 159 \\
Osaka Kyoiku, Japan           & 51 cm         & R             & 1167 \\
Perugia, Italy                & 40 cm         & V, R, I       & 140 \\
Roque (KVA), Canary Islands   & 35 cm         & B, V, R       & 653 \\
Roque (NOT), Canary Islands   & 256 cm        & U, B, V, R, I & 52 \\
Sabadell, Spain               & 50 cm         & B, R          & 4 \\
San Pedro Mart\'\i r, Mexico  & 150 cm        & B, V, R, I    & 185 \\
Shanghai, China               & 156 cm        & V, R          & 36 \\
Skinakas, Crete               & 130 cm        & B, V, R, I    & 156 \\
Sobaeksan, Korea              & 61 cm         & B, V, R, I    & 133 \\
St. Louis, Missouri, USA      & 35 cm         & B, R          & 16 \\
Torino, Italy                 & 105 cm REOSC  & B, V, R, I    & 227 \\
Tuorla, Finland               & 103 cm         & B, V, R       & 1032 \\
\noalign{\smallskip\hrule\smallskip}
\multispan 4 \hss \bf X-Ray Observatory \hss \\
\noalign{\smallskip\hrule\smallskip}
RXTE & PCA & 3 -- 25 keV & 26 \\
\noalign{\smallskip\hrule\smallskip}
\multispan 4 \hss \bf $\gamma$-Ray Observatories \hss \\
\noalign{\smallskip\hrule\smallskip}
STACEE  & Solar-Tower Cerenkov Array  & $> 100$~GeV & 85 \\
VERITAS & Whipple 10 m                & $> 390$~GeV & 31 \\
\enddata
\label{observatories}
\end{deluxetable}

\subsection{\label{radio}Radio observations}

At radio frequencies, the object was monitored using the University 
of Michigan Radio Astronomy Observatory (UMRAO) 26~m telescope, 
at 4.8, 8, and 14.5~GHz, the 14~m Mets\"ahovi Radio Telescope 
of the Helsinki University of Technology, at 22 and 37~GHz, and the 
576-m ring telescope (RATAN-600) of the Russian Academy of Sciences, 
at 2.3, 3.9, 7.7, 11, and 22~GHz. 

At the UMRAO, the source was monitored in the course of the
on-going long-term blazar monitoring program. The data were
analyzed following the standard procedure described in
\cite{aller85}. As mentioned above, the sampling at the lower 
frequencies (4.5 and 8~GHz) was rather poor (about once every 
1 -- 2 weeks) and some individual errors were rather large due to 
source confusion problems with 3C~66B. At 14.5~GHz, a slightly 
better sampling of $\sim 1$ -- 2 observations per week, at least
until the end of 2003, was achieved with relative flux errors
of typically a few percent. The resulting 14.5~GHz light curve 
is included in Fig.~\ref{timeline}. It seems to indicate generally
moderate variability ($\Delta F / F \lesssim 25$~\%) on time 
scales of $\gtrsim 1$~week, though a discrete auto-correlation
analysis \citep{ek88} does not reveal any significant structure, 
primarily due to the insufficient sampling. 

At 22 and 37~GHz, 3C~66A has been monitored using the 14~m radio
telescope of the Mets\"ahovi Radio Observatory of the Helsinki University
of Technology. The data have been reduced with the standard procedure 
described in \cite{ter98}. The 22 and 37~GHz light curves reveal
moderate-amplitude ($\Delta F / F \lesssim 30$~\%), erratic variability 
which is clearly under-sampled by the available data. The discrete
auto-correlation functions indicate a time scale of a few days for
this short-term variability. 

We point out that the observed erratic variability may, at least in 
part, be a consequence of interstellar scintillation. At the Galactic
coordinates of the source, the transition frequency, where the interstellar
scattering strength (a measure of the phase change induced by interstellar
scattering) becomes unity, is $\sim 7$~GHz \citep{walker98}. At higher
frequencies, as considered here, scattering will occur in the weak
scattering regime, with only small phase changes through interstellar
scattering. In this regime, we find the fractional rms variability amplitude 
for a point source due to interstellar scintillation as 0.36 and 0.09 at a 
frequency of 14.5 and 37~GHz, respectively, and the respective variability 
time scales due to interstellar scintillation are 1.4 and 0.9~hours.
Consequently, in particular at lower frequencies ($\lesssim 20$~GHz),
a substantial fraction of the observed variability may well be due to 
interstellar scintillation.

The RATAN-600 has monitored 3C~66A, performing 17 observations from
mid-1997 through Oct. 2003. The last two of these observations (Oct. 
11 and 14) coincided with our core campaign, and provided additional 
frequency coverage at 2.3, 3.9, 7.7, 11, and 22~GHz. Those data were 
analyzed as described in detail in \cite{kovalev99}, and the time 
average of the resulting fluxes from the two observations in Oct. 
2003 are included in the spectral energy distribution shown in 
fig. \ref{sed_plot}.

\begin{figure}
\plotone{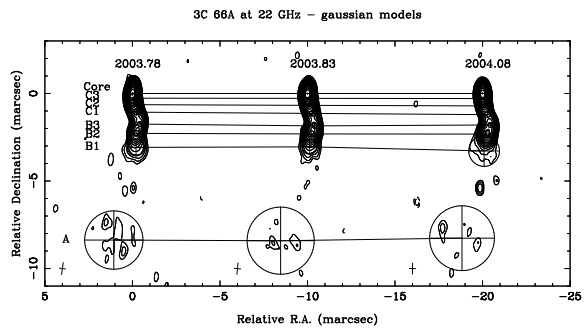}
\caption{22-GHz VLBA maps of 3C~66A during the three epochs in 2003/2004
contemporaneous with our campaign. Overlaid are the fit results from 
decomposing the structure into a sum of Gaussian components.}
\label{VLBA_22_image}
\end{figure}

A total of 9 observing epochs using the VLBA\footnote{The VLBA is a facility
of the National Radio Astronomy Observatory (NRAO). The NRAO is a facility
of the National Science Foundation, operated under cooperative agreement
by Associated Universities, Inc.} 
have been approved to accompany our multiwavelength campaign on 3C~66A. 
In this paper, we describe details of the data analysis and results of 
the first (2003.78; VLBA observing code BS133A), the second (2003.83; 
BS133B), and the fourth (2004.08; BS133E) epoch, concentrating on the 
results at 22 and 43~GHz. The third epoch (2003.96) suffered from radio 
frequency interference, and its reduction is in progress. Results of 
the complete set of all 9 VLBA observing epochs, including all 6 
frequencies (2.3, 5, 8.4, 22, 43, and 86~GHz) together with polarization 
data,  will be presented in a separate paper (Savolainen et al., 2005, 
in preparation).

\begin{figure}
\plotone{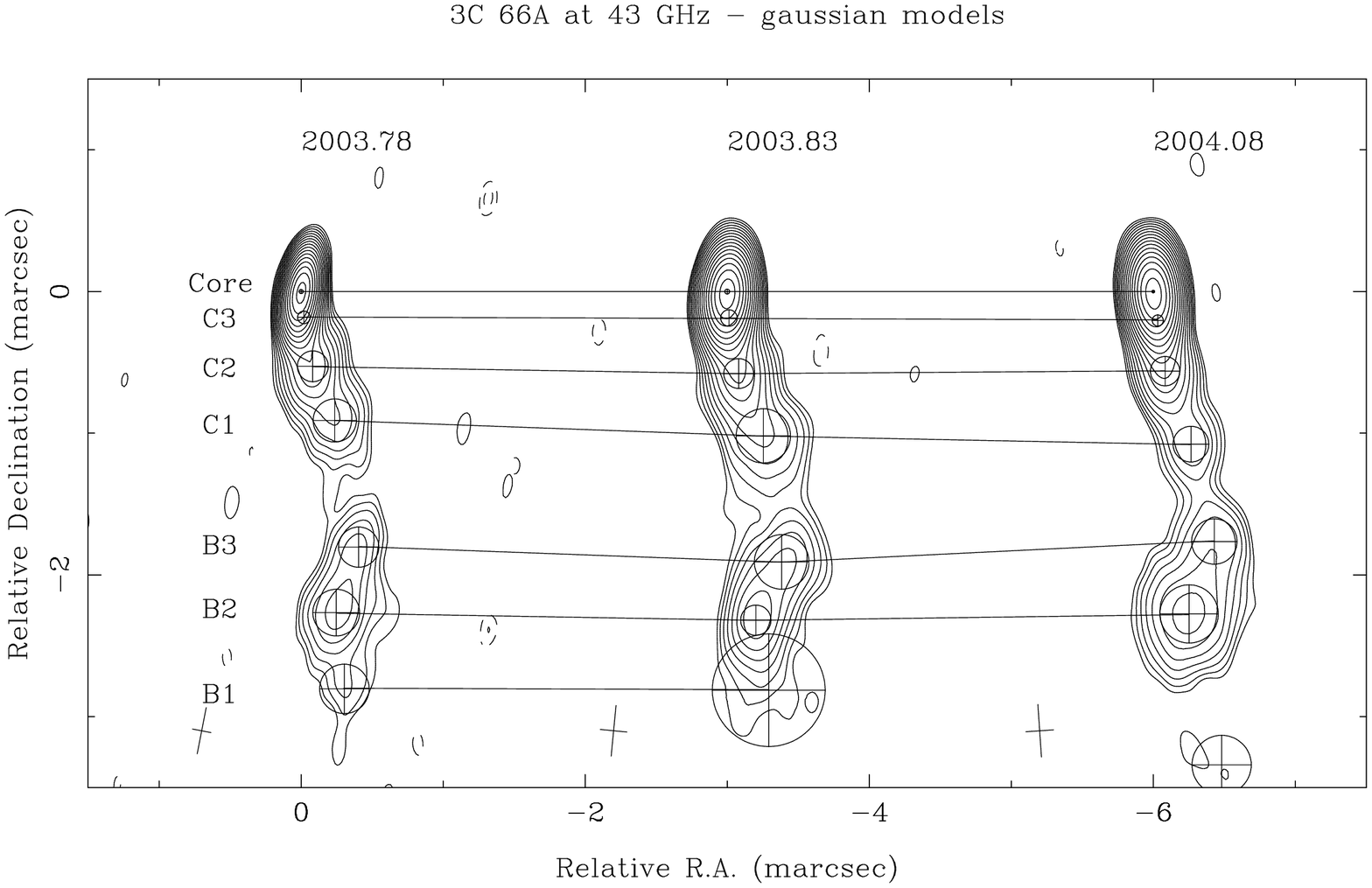}
\caption{43-GHz VLBA maps of 3C~66A during the three epochs in 2003/2004
contemporaneous with our campaign. Overlaid are the fit results from 
decomposing the structure into a sum of Gaussian components.}
\label{VLBA_43_image}
\end{figure}

The data were correlated on the VLBA correlator and were postprocessed 
with the NRAO Astronomical Image Processing System, AIPS \citep{aips}, 
and the Caltech DIFMAP package \citep{difmap}. Standard methods for 
VLBI data reduction and imaging were used. {\it A priori} amplitude 
calibration was performed using measured system temperatures and 
gain curves in the AIPS task APCAL. At this point, a correction for
atmospheric opacity was also applied. After removal of the parallactic angle
phase, single-band delay and phase offsets were calculated manually by 
fringe-fitting a short scan of data of a bright calibrator source (0420-014). 
We did manual phase calibration instead of using pulse-cal tones, because there 
were unexpected jumps in the phases of the pulse-cal during the observations.
Global fringe-fitting was performed with the AIPS task FRING using Los
Alamos (LA) as a reference antenna. The AIPS procedure CRSFRING was used to
remove the delay difference between right- and left-hand systems (this is
needed for polarization calibration). A bandpass correction was determined and
applied before averaging across the channels, after which the data were
imported into DIFMAP.

In DIFMAP, the data were first phase self-calibrated using a point source model,
and then averaged in time. We performed data editing in a station-based manner,
and ran several iterations of CLEAN and phase self-calibration in Stokes I.
After a reasonable fit to the closure phases was obtained, we also performed
amplitude self-calibration, first with a solution interval corresponding to
the whole observation length. The solution interval was then gradually shortened
as the model improved by further CLEANing. Final images were produced using the
Perl library FITSplot\footnote{\tt http://personal.denison.edu/$\sim$homand/}. For 
final images, normal weighting of the uv-data was used in order to reveal the 
extended, low surface brightness emission.

We have checked the amplitude calibration by comparing the
extrapolated zero baseline flux density of the compact source 0420-014
to the single-dish flux measurements at Mets\"ahovi. The fluxes are
comparable: at 22~GHz the VLBI flux is on average 5~\% lower than the
single-dish flux, and the 43~GHz flux is on average 8~\% lower than the
nearly simultaneous single-dish measurements at 37~GHz. The small
amount of missing flux is probably due to some extended emission
resolved out in the high frequency VLBI images. The integrated VLBI
flux of 3C~66A is about 30~\% smaller than the corresponding single-dish
value both at 22~GHz and at 43~GHz, but since the source has notable
kpc-scale structure evident in the VLA images \citep{price93,taylor96}, 
it is most likely that the missing flux comes from this kpc-scale jet 
which is resolved out in our VLBI images. Thus, we conclude that the 
accuracy of the amplitude calibration is better than 10~\% at both 
frequencies.

\begin{deluxetable}{ccc}
\tabletypesize{\scriptsize}
\tablecaption{Core brightness temperatures and corresponding lower limits,
calculated from the measured 43~GHz VLBA fluxes and component sizes}
\tablewidth{0pt}
\tablehead{
\colhead{Epoch} & \colhead{$T_b^{\rm core}$ [K]} & 
\colhead{$T_{\rm b, min}^{\rm core}$ [K]}
}
\startdata
2003.78 & $2.7 \times 10^{11}$ & $1.1 \times 10^{11}$ \\
2003.83 & $2.0 \times 10^{11}$ & $1.0 \times 10^{11}$ \\
2004.08 & $6.6 \times 10^{11}$ & $1.4 \times 10^{11}$ \\
\enddata
\label{core_Tb}
\end{deluxetable}

In order to estimate the parameters of the emission regions in the jet, we did
model fitting to the self-calibrated visibilities in DIFMAP. The data were
fitted with circular Gaussian model components and we sought to obtain the
best possible fit to the visibilities and to the closure phases. Several
starting points were tried in order to avoid a local minimum fit. The 
results of this fitting procedure are included in Figs. \ref{VLBA_22_image}
and \ref{VLBA_43_image}. In all the model-fits, the core of the VLBI jet 
is the northernmost and also the brightest component. Beyond the core, 
we have divided the jet into three regions named A, B and C. Closest 
to the core is the region C, which consists of three components with 
decreasing surface brightness as a function of the distance from the 
core. The region B, also made of three components, shows a clear bending 
of the jet together with re-brightening. The observed increase in the 
flux at this point could plausibly be due to increased Doppler boosting 
caused by the bending of the jet towards our line of sight. The  region 
A shows weak extended emission at 8 mas from the core at 22~GHz. This 
emission is more pronounced at lower frequencies.

\begin{deluxetable}{ccccccc}
\tabletypesize{\scriptsize}
\tablecaption{VLBA component fluxes, core distances (Rad.), component position
angles (PA), and circular component diameters (Size)}
\tablewidth{0pt}
\tablehead{
\colhead{Epoch} & \colhead{Freq. [GHz]} & \colhead{Comp. ID} & \colhead{Flux [mJy]} &
\colhead{Rad. [mas]} & \colhead{PA [deg]} & \colhead{Size [mas]}
}
\startdata
2003.78 &    22 GHZ     & Core   &      387     &     0.0    &     0.0    &   0.047\\
	&		& C3     &       61     &     0.295  &     -172.9 &   0.136\\
        &		& C2     &       36     &     0.658  &     -168.2 &   0.164\\
        &		& C1     &       19     &     1.096  &     -167.9 &   0.309\\
        &		& B3     &       21     &     1.803  &     -166.1 &   0.278\\
        &		& B2     &       40     &     2.287  &     -174.6 &   0.331\\
        &		& B1     &       19     &     3.086  &     -176.2 &   0.970\\
        &		& A      &       37     &     8.439  &      172.8 &   3.333\\
\\
	&      43 GHz   & Core   &     355      &     0.0    &        0.0 &   0.029\\
        &		&   C3   &      63      &     0.183  &     -174.0 &   0.089\\
        &		&   C2   &      43      &     0.534  &     -171.1 &   0.218\\
        &		&   C1   &      24      &     0.940  &     -165.4 &   0.304\\
        &		&   B3   &      18      &     1.849  &     -167.3 &   0.284\\
        &		&   B2   &      30      &     2.279  &     -173.8 &   0.329\\
        &		&   B1   &      9       &     2.821  &     -173.8 &   0.349\\
\\
2003.83 &  22 GHz  	& Core   &     375      &     0.0    &     0.0    &   0.032\\
        &		&   C3   &      59      &     0.282  &     -172.4 &   0.118\\
        &		&   C2   &      38      &     0.685  &     -170.2 &   0.193\\
        &		&   C1   &      18      &     1.167  &     -167.3 &   0.298\\
        &		&   B3   &      23      &     1.892  &     -167.7 &   0.274\\
        &		&   B2   &      39      &     2.320  &     -174.8 &   0.358\\
        &		&   B1   &      13      &     3.051  &     -177.6 &   0.836\\
        &		&   A    &      41      &     8.546  &      169.7 &   3.836\\
\\
        &  43 GHz	&   Core &     369      &     0.0    &      0.0   &   0.034\\
        &		&   C3   &      76      &     0.187  &     -176.3 &   0.116\\
        &		&   C2   &      34      &     0.584  &     -172.1 &   0.210\\
        &		&   C1   &      26      &     1.050  &     -166.0 &   0.383\\
        &		&   B3   &      26      &     1.947  &     -168.6 &   0.385\\
        &		&   B2   &      20      &     2.328  &     -175.0 &   0.213\\
        &		&   B1   &      16      &     2.829  &     -174.1 &   0.795\\
\\
2004.08 &  22 GHz 	&   Core &     363      &     0.0    &      0.0   &   0.046\\
        &		&   C3   &      64      &     0.289  &     -172.4 &   0.132\\
        &		&   C2   &      32      &     0.712  &     -172.0 &   0.235\\
        &		&   C1   &      12      &     1.241  &     -167.2 &   0.311\\
        &		&   B3   &      22      &     1.860  &     -165.7 &   0.296\\
        &		&   B2   &      45      &     2.333  &     -174.2 &   0.386\\
        &		&   B1   &      18      &     3.285  &     -178.5 &   1.754\\
        &		&    A   &      38      &     8.348  &      171.9 &   3.702\\
\\
    	&  43 GHz 	&   Core &     372      &     0.0    &      0.0   &   0.019\\
        &		&   C3   &      59      &     0.207  &     -171.2 &   0.078\\
        &		&   C2   &      30      &     0.568  &     -171.6 &   0.207\\
        &		&   C1   &      13      &     1.109  &     -166.2 &   0.252\\
        &		&   B3   &      15      &     1.816  &     -166.3 &   0.320\\
        &		&   B2   &      40      &     2.289  &     -173.7 &   0.411\\
        &		&   X1   &      4       &     3.375  &     -171.8 &   0.417\\
\enddata
\label{vlba_comp_tab}
\end{deluxetable}

\begin{figure}
\plotone{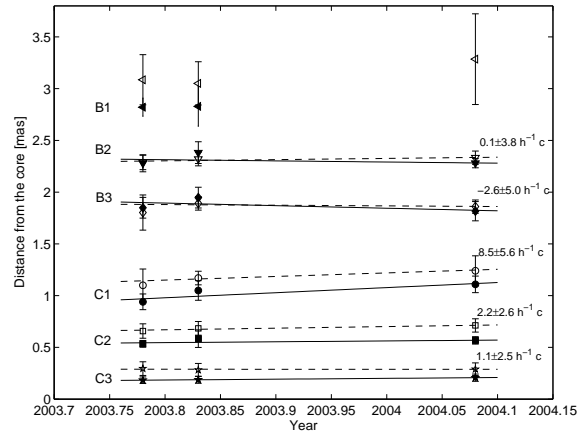}
\caption{
Separation from the core of the Gaussian fit components shown in 
Figs. \ref{VLBA_22_image} and \ref{VLBA_43_image}, as a function of time.
Filled symbols refer to 22~GHz, open symbols to 43~GHz. The solid (dashed) 
lines show linear least-squares fits to the positions of 22~GHz (43~GHz)
components. The velocity results are the average of the results from the
two frequencies. All components except C1 are consistent with being 
stationary.}
\label{slmotion43}
\end{figure}

Fig. \ref{slmotion43} shows the separation of the model components from 
the core, at each epoch. The components A - C are weak compared to the
core (see Table \ref{vlba_comp_tab}), which renders the estimation of errors 
of the model parameters using programs like Difwrap \citep{lovell00} problematic. 
Thus, in Fig. \ref{slmotion43}, we have assumed the following uncertainties in the
component position accuracy: 1) for bright features with $SNR > 50$, the
positions are accurate to 20~\% of the projection of the elliptical beam
FWHM onto a line joining the center of the core to the center of the
components; 2) for weaker components with $SNR < 50$, uncertainties are
50~\% of the beam size or 50~\% of the size of the knot --- whichever is
larger. These estimates are based on the experience with other sources
(Savolainen et al. 2005, in preparation; Wiik et al. 2005, in
preparation), and they follow the results by \cite{jorstad05},
where uncertainties of model-fit parameters for a large number of
observations are estimated. Positional errors of similar size are also
reported by \cite{homan01}, who have estimated the uncertainties
from the variance of the component position about the best-fit
polynomial.

There is a $\sim 0.1$~mas shift of the components C1 -- C3 between 
the two frequencies, with the 22~GHz model components appearing further 
downstream. This cannot be explained by an opacity effect, since in that 
case the components would have shifted in the opposite direction: at
higher frequency we would expect to see emission from the region
closer to the apex of the jet than at lower frequency, if the opacity
effect is significant. This obviously is not the case in Fig. 
\ref{slmotion43}, and moreover, for the components B2 and B3 the 
positions at 22 and 43~GHz are coincident. However, there is a simple 
explanation for the observed shift, if the brightness profile along 
the jet in region C is smooth. Namely, the flux in the outer part
of the region C decreases more steeply at 43~GHz than at 22~GHz (see
Figs. \ref{VLBA_22_image} and \ref{VLBA_43_image}). As the region 
C is modelled by the same number of components at both frequencies, 
the model-fit procedure shifts the 22~GHz components downstream 
relative to 43~GHz components in order to represent the power-law.

Fig. \ref{slmotion43} shows that, except for
one component, all the components revealed by the analysis of the three 
epochs considered here are consistent with zero proper motion. Although 
a monitoring effort over a longer time scale will be necessary to settle this 
issue, our analysis shows slower component
motion than what is presented by \cite{jorstad01}, where fast
superluminal motions of several components, with speeds ranging from 
$9 \, h^{-1}$~c to $19 \, h^{-1}$~c were found. Note, however, that the 
data used by \cite{jorstad01} covered a much longer time frame and refers 
to different components and a rather different state of activity of 3C~66A. 
Moreover, there are also similarities between the data sets:
\cite{jorstad01} reported a stationary component (their component
"C") at distance of $\sim 0.5$~mas from the core, which could correspond
to our component C2. New 7~mm VLBA monitoring data presented in a
recent paper by \cite{jorstad05} suggest that there are two kinds
of components in 3C~66A: fast and very weak components with apparent
speeds $> 20$~c, and stronger components with moderate velocities of
1.5 -- 5~c. The components C1 -- C3 in our analysis could be qualitatively
similar to the latter. More solid 
conclusions concerning the question of superluminal motions of individual
components might be possible after the final analysis of all 9 epochs of 
the entire VLBA monitoring program of 3C66A proposed in connection with
this campaign (Savolainen et al., 2005, in preparation). Based on the
analysis of the three epochs contemporaneous with the other observations 
of this campaign, one component, C1, shows superluminal motion of $(8.5 
\pm 5.6) \, h^{-1}$~c = $(12.1 \pm 8.0)$~c for $h = 0.7$.

The essentially zero proper motion of most of the components together with
the observed trend of the brightness temperature vs. distance along the jet 
(see \S \ref{VLBA_parameters}) suggest that in the case of 3C~66A the VLBA 
model-fit components might not correspond  to actual 'knots' in the jet, 
but a rather smooth flow with simple brightness profile (such as a power 
law) might better describe the jet during this campaign --- at least in 
region C. More detailed parameter estimates will be extracted from these 
results in \S \ref{VLBA_parameters}.

\subsection{\label{infrared}Infrared observations}

In the context of the extensive WEBT campaign (see \S \ref{optical}), 
3C~66A was also observed at near-infrared wavelengths in the J, H, and K/K'
bands at three observatories: The Campo Imperatore 1.1-m telescope of the
Infrared Observatory of the Astronomical Observatory of Rome, Italy, the 
1.2-m telescope (using the NICMOS-3 HgCdTe IR array with 256 $\times$ 256
pixels, with each pixel corresponding to 0.96" on the sky) at the Mt. Abu 
Infrared Observatory (MIRO) at Mt. Abu, India, and at the 2.56-m Nordic 
Optical Telescope (NOT) on Roque de los Muchachos on the Canary Island 
of La Palma. The primary data were analyzed using the same standard 
technique as the optical data (see \S \ref{optical}), including 
flat-field subtraction, extraction of instrumental magnitudes, calibration 
against comparison stars to obtain standard magnitudes, and de-reddening. The 
sampling was generally not dense enough to allow an improvement of the data 
quality by re-binning. Unfortunately, the three observatories did not perform 
any observations on the same day, so that the cross-calibration between 
different instruments is problematic. We have therefore opted not to correct 
for possible systematic offsets beyond the calibration to standard magnitudes 
of well-calibrated comparison stars (see \S \ref{optical}). 

\begin{figure}
\plotone{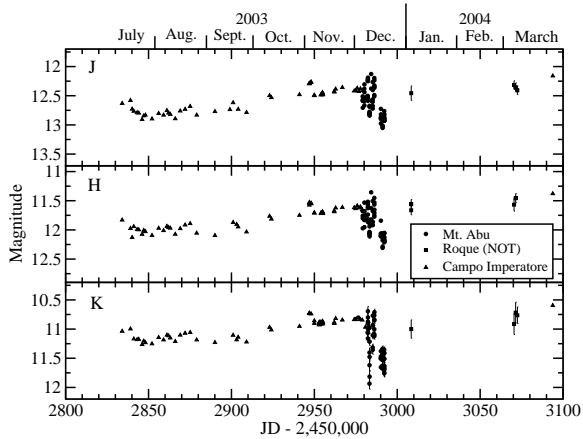}
\caption{Light curves in the near-infrared J, H, and K bands over the
entire duration of the campaign.}
\label{IR_lc}
\end{figure}

The resulting H-band light curve is included in Fig. \ref{timeline}, and
a comparison of all three IR band light curves (J, H, K/K') is shown in 
Fig. \ref{IR_lc}. Generally, we find moderate variability of $\Delta 
(J, H, K/K') \lesssim 0.2$ within $\sim$~a few days. The various infrared 
bands trace each other very well, and there does not appear to be any
significant time lags between the different bands. It is obvious that
there are issues with the cross calibration between the three observatories.
In particular, the Mt. Abu observations (using the J, H, and K' filters) 
in mid-December of 2003 indicate a significant drop of the average level 
of near-infrared flux. While there appears to be a similar flux drop 
around the same time in the radio light curves, the much more densely 
sampled optical light curves do not indicate a similar feature. However, 
it is obvious that the near-IR data around this time exhibit an unusually 
large scatter (by $\sim 0.5$~mag on sub-hour time scales), they may be 
affected by calibration uncertainties. We note that rebinning of the IR 
data in time bins of up to 30 minutes did not improve the quality of the
data because of the extreme nature of the apparent flux fluctuations. We
also point out that the use of the K' filter at Mt. Abu vs. the K filter
at the other two IR observatories, could only explain offsets of less than
the individual measurement errors, so this effect was neglected in our
analysis.

\subsection{\label{optical}Optical observations}

In the optical component of the extensive WEBT campaign, 24
observatories in 15 countries contributed 7611 individual photometric
data points. The observing strategy and data analysis followed to a 
large extent the standard procedure described previously for a similar
campaign on BL~Lacertae in 2000 \citep{villata02}. For more information
about WEBT campaigns see also: 
\cite{villata00,villata02,villata04a,villata04b,raiteri01,raiteri05}

It had been suggested that, optimally, observers perform photometric 
observations alternately in the B and R bands, and include complete 
(U)BVRI sequences at the beginning and the end of each observing
run. Exposure times should be chosen to obtain an optimal compromise
between high precision (instrumental errors less than $\sim 0.03$~mag
for small telescopes and $\sim 0.01$~mag for larger ones) and high time
resolution. If this precision requirement leads to gaps of 15 -- 20
minutes in each light curve, we suggested to carry out observations 
in the R band only. Observers were asked to perform bias (dark) 
corrections as well as flat-fielding on their frames, 
and obtain instrumental magnitudes, applying either aperture
photometry (using IRAF or CCDPHOT) or Gaussian fitting for the
source 3C~66A and the comparison stars no. 13, 14, 21, and 23 in the
tables of \cite{gp01}, where high-precision standard magnitudes for
these stars have been published. This calibration has then been used
to convert instrumental to standard photometric magnitudes for each
data set. In the next step, unreliable data points (with large error
bars at times when higher-quality data points were available) were
discarded. Our data did not provide evidence for significant variability
on sub-hour time scales. Consequently, error bars on individual data
sets could be further reduced by re-binning on time scales of typically
15 -- 20~min. 

Finally, there may be systematic offsets between different instruments 
and telescopes. Wherever our data sets contained sufficient independent
measurements to clearly identify such offsets, individual data sets 
were corrected by applying appropriate correction factors. In the case
of BL~Lacertae and similar sources \citep[e.g.,][]{villata02}, such 
corrections need to be done on a night-by-night basis since changes in
the seeing conditions affect the point spread function and thus the
precise amount of the host galaxy contamination. However, in the absence
of a significant host galaxy contribution, these systematic effects 
should be purely instrumental in nature and should thus not depend 
on daily seeing conditions. Thus, we opted to introduce only global 
correction factors for entire single-instrument data sets on 3C~66A. 
This provided satisfactory results without obvious residual 
inter-instrumental inconsistencies. The resulting offsets are 
listed in Tab. \ref{offsets}.

\begin{figure}
\plotone{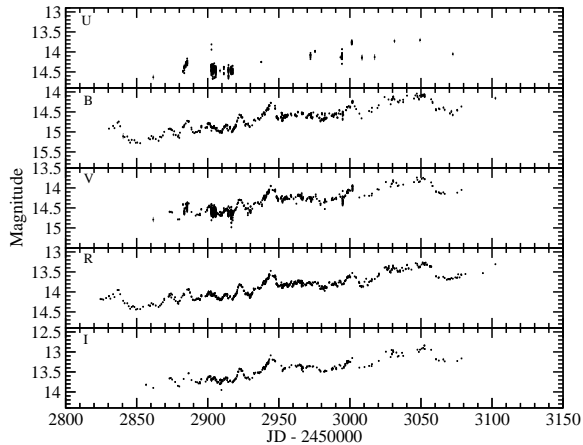}
\caption{Light curves in the optical U, B, V, R, and I bands over the
entire duration of the campaign.}
\label{optical_lc}
\end{figure}

In order to provide information on the intrinsic broadband 
spectral shape (and, in particular, a reliable extraction 
of B - R color indices), the data were  de-reddened using 
the Galactic Extinction coefficients of \cite{schlegel98}, 
based on $A_B = 0.363$~mag and $E(B-V) = 
0.084$~mag\footnote{\tt http://nedwww.ipac.caltech.edu/}. As 
mentioned in the introduction, the R magnitude of the host 
galaxy of 3C~66A is $\sim 19$~mag, so its contribution is 
negligible, and no host-galaxy correction was applied.

\begin{deluxetable}{ccccc}
\tabletypesize{\scriptsize}
\tablecaption{Inter-instrumental correction offsets for optical data}
\tablewidth{0pt}
\tablehead{
\colhead{Observatory} & \colhead{B} & \colhead{V} & \colhead{R} & \colhead{I} 
}
\startdata
Armenzano&	       +.02&    -&     +.02&	-\\
Bell&			-&	-&	-&	-\\
Boltwood&	       -.02&	-&	-&	-\\
Catania&		-&	-&	-&	-\\
Crimea&			-&	+.03&	-&	-\\
Heidelberg&		-&	-&	-&	-\\
MDM& 			-&	-&	-&	-\\
Michael Adrian&		-&	-&	-&	-\\	
Mt. Lemmon&		-&	-&	-&	-\\
Mt. Maidanak&	       +.01&	-&	-&	-\\
Nyrola&			-&	-&     -.03&	-\\
Osaka&			-&	-&	-&	-\\
Perugia&		-&	+.08&     +.05&   +.04\\
Roque(NOT)&		-&	-&	-&	-\\
Sabadell&		-&	-&	-&	-\\
San Pedro Martir&      +.08&	-&     +.07&	-\\
Shanghai&	        -&      -&     -.03&	-\\
Skinakas&	       -.01&   -.04&   -.04&   -.05\\
Sobaeksan&	        -&      -&      -&      -\\
St. Louis& 	       -.06&	-&	-&	-\\
Torino&	               -.09&    -&      -&     -.03\\
Tuorla&  	        -&      +.02&     -.02&    -\\
\enddata
\label{offsets}
\end{deluxetable}

\subsubsection{\label{opt_lightcurves}Light curves}

\begin{figure}
\plotone{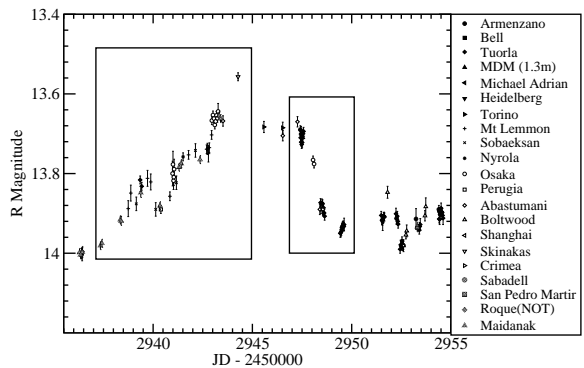}
\caption{Details of the R-band light curve during the major outburst around
Nov. 1, 2003. The boxes indicate the light curve segments shown in more 
detail in Figs. \ref{nov1_rise} -- \ref{nov1_decay}.}
\label{nov1_total}
\end{figure}

\begin{figure}
\plotone{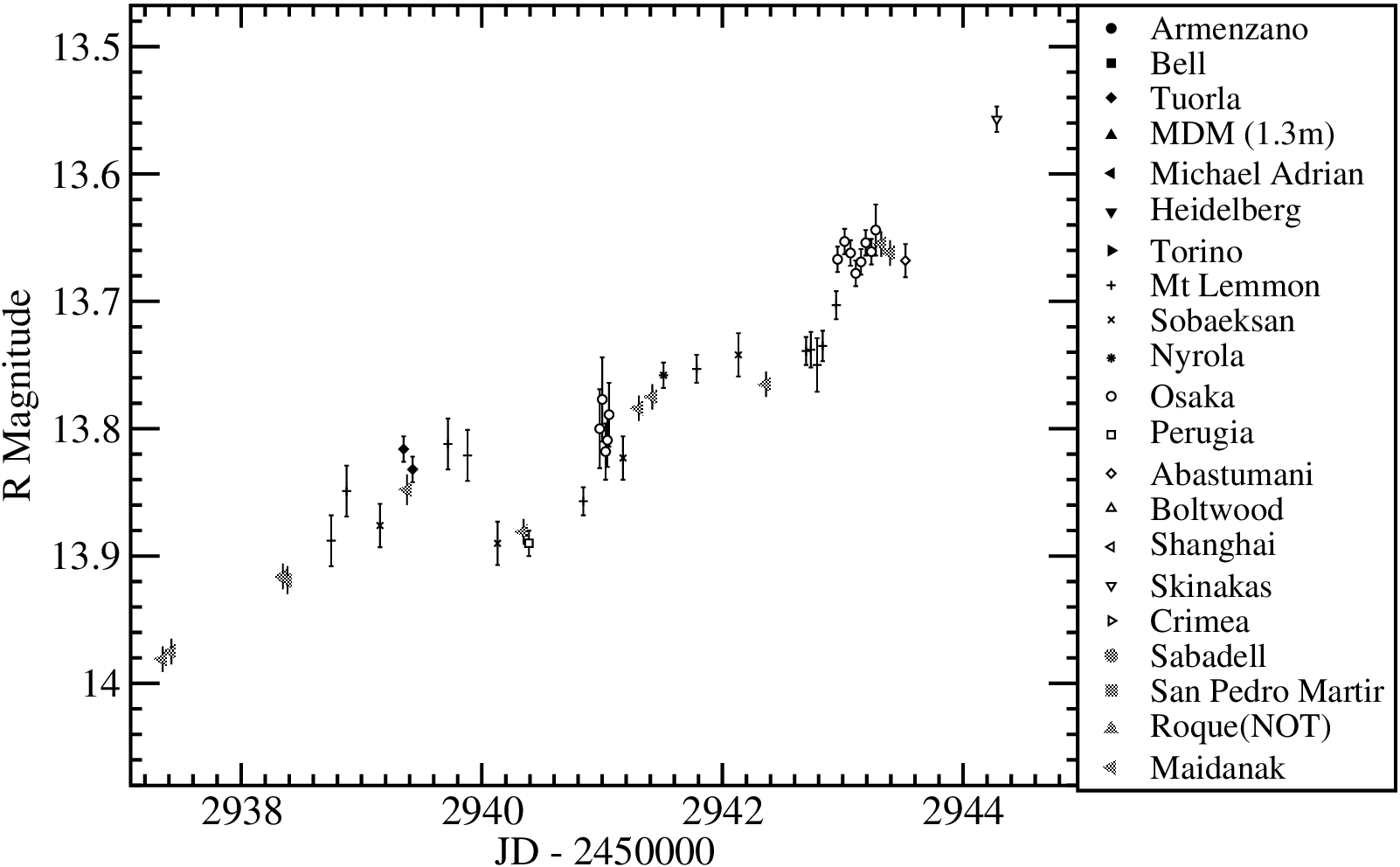}
\caption{Details of the R-band light curve during the rising phase of the
major outburst around Nov. 1, 2003. 
}
\label{nov1_rise}
\end{figure}

\begin{figure}
\plotone{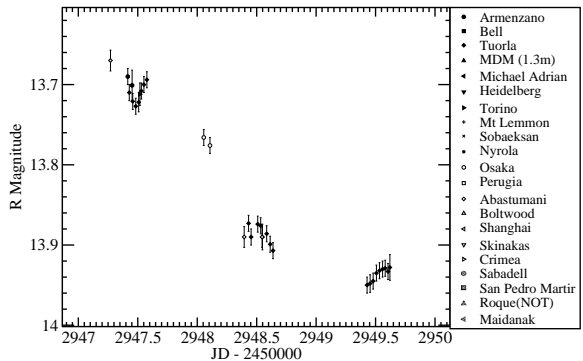}
\caption{Details of the R-band light curve during the decaying phase of the
major outburst around Nov. 1, 2003. 
}
\label{nov1_decay}
\end{figure}

\begin{figure}
\plotone{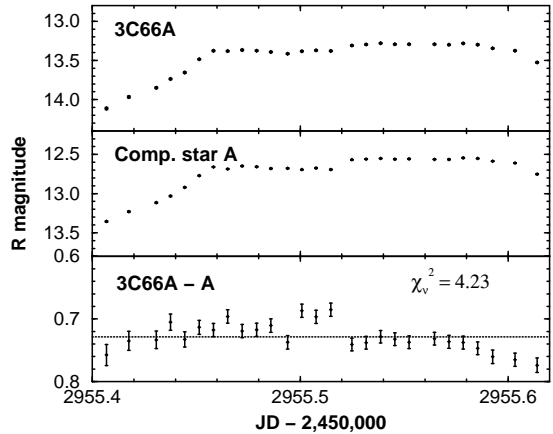}
\caption{Instrumental magnitudes of 3C~66A (top panel) and comparison star A 
(central panel) of the University of Torino data of JD~2,452,955 (= Nov. 11, 2003)
The difference (bottom panel) shows clear evidence for intranight variability
during the central portion of this observation ($\sim$~JD~2,452,955.4
-- $\sim$~JD~2,452,955.5).}
\label{intradayvar}
\end{figure}

As a consequence of the observing strategy described above,
the R- and B-band light curves are the most densely sampled 
ones, resulting in several well-sampled light curve segments 
over $\sim 5$ -- 10~days each, with no major interruptions,
except for a gap of a few hours due to the lack of coverage 
when 3C~66A would have been optimally observable from locations 
in the Pacific Ocean. The R-band light curve 
over the entire duration of the campaign is included in 
Fig.~\ref{timeline} and compared to the light curves at all 
other optical bands in Fig.~\ref{optical_lc}. These figures
illustrate that the object underwent a gradual brightening
throughout the period July 2003 -- February 2004, reaching
a maximum at $R \approx 13.4$ on Feb. 18, 2004, followed by
a sharp decline by $\Delta R \sim 0.4$~mag within $\sim 15$~days.
On top of this overall brightening trend, several major outbursts
by $\Delta R \sim 0.3$ -- $0.5$~mag on time scales of $\sim 10$~days
occurred. The two most dramatic ones of these outbursts peaked on 
July 18, 2003, and Nov. 1, 2003. Details of the Nov. 1 outburst 
are displayed in Figs. \ref{nov1_total} -- \ref{nov1_decay}, 
respectively. We find evidence for intraday microvariability of 
$\Delta R \sim 0.05$~mag on time scales down to $\sim 2$~hr. 
One example for such evidence is illustrated in Fig. \ref{intradayvar} 
which shows the intranight data from the Torino Observatory for
JD~2,452,955 (= Nov. 11, 2003). The figure presents the instrumental
magnitudes of 3C~66A, comparison star A, and the difference of
both. While the difference seems to be affected by seeing
effects at the beginning and end of the observation (rising and
falling in tandem with the instrumental magnitudes), there is
evidence for intraday variability around $\sim$~JD~2,452,955.5, 
where there were only very minor changes in the atmospheric opacity.

Visual inspection of individual major outbursts suggests periods 
of more rapid decline than rise. 

\subsubsection{\label{periodicity}Periodicity analysis}

For several blazars, including 3C~66A, periodicities on various
time scales, from several tens of days, to several years, have been 
claimed \citep[see, e.g.][and references therein for a more complete
discussion]{rieger04}. In order to search for a possible periodicity 
in the optical R-band light curve from our WEBT campaign, we
have performed a Fourier analysis of the R-band light curve after 
interpolation of the light curve over the (short) gaps in the 
available data. Such an analysis becomes unreliable for periods
of more than $\sim 1/10$ of the length of the data segment. 
Consequently, our data can be used for a meaningful analysis on 
time scales of $\tau \lesssim 30$ -- 50~d. No evidence for a periodicity 
in this range has been found. However, we note a sequence of large
outbursts on July 18 (MJD 52838), Sept. 5 (MJD 52887), Nov. 1
(MJD 52944), Dec. 28 (MJD 53001), 2003, and Feb. 18, 2004 (MJD 53053),
which are separated by intervals of 49, 57, 57, and 52 days, respectively.
These intervals appear remarkably regular, and only slightly shorter
than the 65-day period claimed by \cite{lainela99}. The duration 
of our monitoring campaign is too short to do a meaningful
analysis of the statistical significance of this quasi-periodicity.
This question will be revisited in future work, on the basis of a 
larger data sample, including archival optical data.

\subsection{\label{xrays}X-ray observations}

3C~66A was observed by the Rossi X-Ray Timing Explorer ({\it RXTE}) 
26 times between September 19 and December 27, 2003, for a total 
observation time of approximately 200~ks. Analysis of the {\it RXTE PCA} 
data was carried out using faint-source procedures as described in the 
{\it RXTE} Cookbook. Standard selection criteria were used to remove 
disruptions from the South Atlantic Anomaly, bright Earth, instrument 
mispointing, and electron contamination. For the entire data set, only 
PCUs 0 and 2 were activated, but data from PCU~0 was discarded due to 
its missing propane veto. There is also a persistent problem in the 
background model of the PCA around the 4.78 keV Xenon L-edge, which 
causes a small anomaly in count rates near this energy. For faint 
sources like 3C~66A, this anomaly can be significant, so photons in 
the three energy bins surrounding this energy were discarded.

\begin{figure}
\begin{center}
\includegraphics[angle=270,scale=0.3]{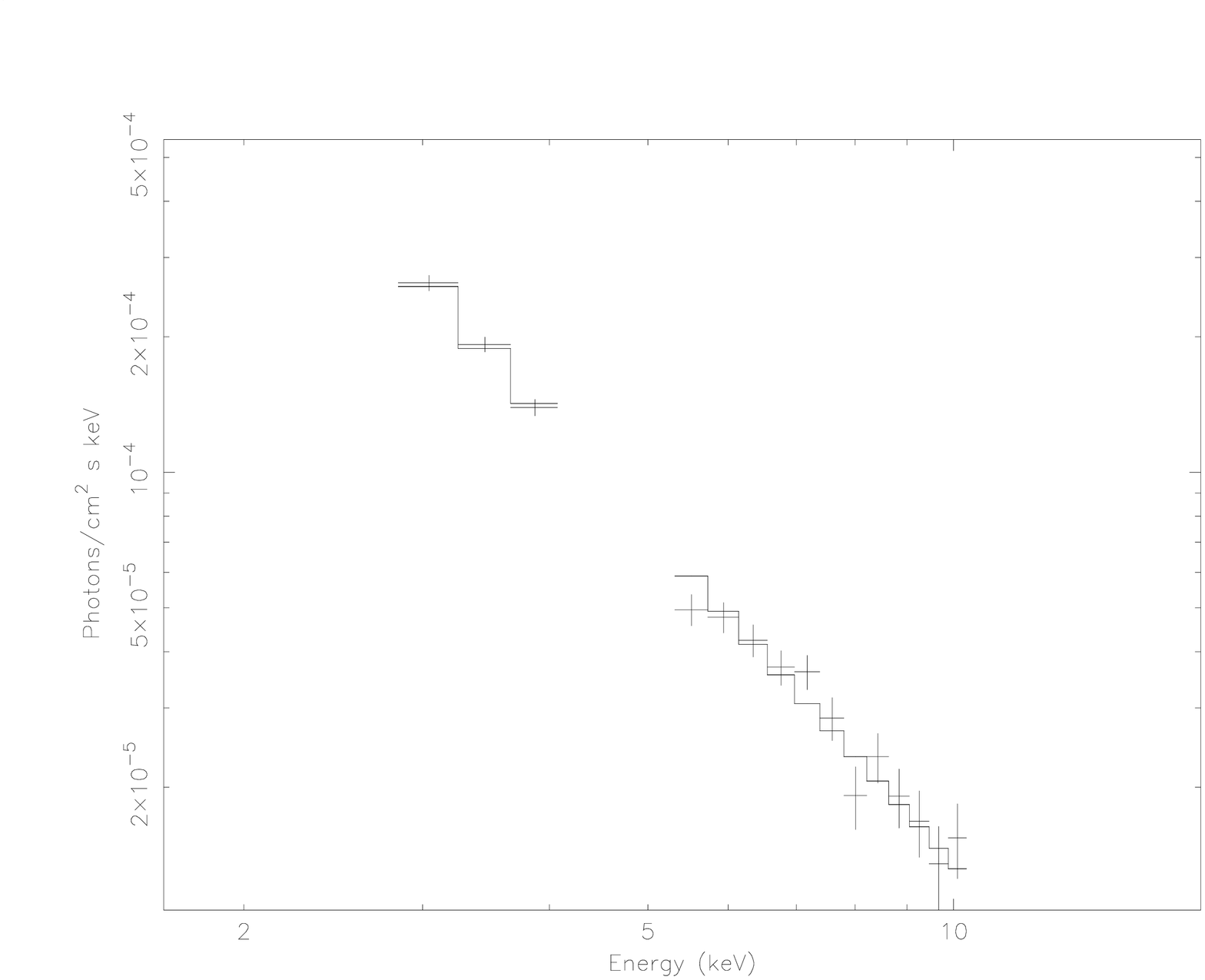}
\end{center}
\caption{Average RXTE spectrum (3 -- 10~keV) and power-law fit for all 
observations of 3C~66A taken as part of the multiwavelength campaign.
See text for best-fit parameters.}
\label{xray_spec}
\end{figure}

Photons from 3 -- 10~keV were extracted from the data using only the top
layer of PCU2. Extracting higher-energy photons proved unreliable, as the
source flux was close to the PCA source-confusion limit. A spectrum was
compiled using data from all 26 source observations. The spectrum was fit
to a simple power-law, which resulted in an energy spectral index of 
$\alpha = 1.47 \pm 0.56$ and a flux $\Phi_{\rm 1 \, keV} = (4.06 \pm 0.49) 
\times 10^{-3}$~photons~cm$^{-2}$~s$^{-1}$~keV$^{-1}$,  with $\chi^{2}/n = 
13.6/14$. Galactic absorption was not included in the fit model, as the 
lack of data below 3 keV precludes a useful constraint on N$_{H}$. A plot 
of the spectrum and fit is shown in Figure \ref{xray_spec}.

The PCA is a non-imaging instrument with a FWHM response of $1^{\circ}$,
so there is the potential that the signal found here is a sum over
multiple sources. Most notably, the FR-1 galaxy 3C~66B lies only
$6^{\prime}$ from 3C~66A and is a known X-ray emitter. Though
indistinguishable from 3C~66A with the PCA, its typical flux is at a 
level of $\sim 2$~\% of the measured PCA flux of 3C~66A \citep{croston03},
and is thus not a significant contributor. In the 3EG catalog, there is
also confusion of 3C~66A with a nearby pulsar, PSR~J0218+4232. At a source
separation of $58^{\prime}$, the PCA response for the pulsar is low for 
3C~66A pointed observations, and thus it does not make a significant
contribution.

The 3 -- 10~keV photons were binned into 24-hour
periods to render a light curve, which is included in Fig. \ref{timeline}. 
No significant flux variations or transient events were seen. To check for 
spectral variations, the data set was divided into four roughly equal 
sub-periods, and power-law spectra fitted to each sub-period. 
Fits to all four sub-periods were consistent with the spectral parameters 
obtained from the entire data set, and all sub-periods were consistent with 
each other. From this we conclude that there were no variations of either 
flux or spectral shape in the X-ray band during the observing period within
the detection capabilities of {\it RXTE}. 

\subsection{\label{gammarays}Gamma-ray observations}

At VHE $\gamma$-rays, 3C~66A was observed contemporaneously with our
broadband campaign by the Solar Tower Air \v Cerenkov Effect Experiment
(STACEE) and by the 10-m Whipple Telescope of the Very Energetic Radiation
Imaging Telescope Array System (VERITAS) collaboration. 

STACEE took a total of 85 28-minute on-off pairs of data, totalling 33.7~hours
of live time on-source. After data-quality cuts and padding, a total of
16.3~hours of on-source live time remained. A net on-source excess of 
1134 events attributed to photons of energy $E > 100$~GeV was seen against
a background of 231,742 events. The details of the STACEE data analysis are 
published in a separate paper \citep{bramel05}. 

\begin{deluxetable}{ccc}
\tabletypesize{\scriptsize}
\tablecaption{99 \% confidence upper limits on the VHE photon flux 
from the STACEE observations for various assumptions of an underlying
intrinsic photon spectral index $\Gamma = \alpha + 1$}
\tablewidth{0pt}
\tablehead{
\colhead{$\Gamma$} & \colhead{$E_{\rm thr}$ [GeV]} & 
\colhead{$dN/dE$ at $E_{\rm thr}$ [photons m$^{-2}$~s$^{-1}$~GeV$^{-1}$]}
}
\startdata
2.0 & 200 & $5.23 \times 10^{-9}$ \\
2.5 & 184 & $9.39 \times 10^{-9}$ \\
3.0 & 150 & $2.26 \times 10^{-8}$ \\
3.5 & 147 & $3.10 \times 10^{-8}$ \\
\enddata
\label{STACEE_UL}
\end{deluxetable}

The source excess quoted above corresponds to a significance of $2.2 \, \sigma$, 
which is insufficient to claim a detection, but can be used to establish firm 
upper flux limits. For various assumptions of an underlying intrinsic source
power-law spectrum, those upper limits are listed in Table \ref{STACEE_UL}.
The data were also binned in 24-hr segments. This revealed no significant
excess on any given day, and thus no evidence for a statistically significant
transient event during the campaign period.

The VERITAS collaboration observed 3C~66A during the period Sept. 27, 2003
-- Jan. 13, 2004, obtaining a total of 31 on-off pairs with typically $\sim
27.6$~min. of on-source live time per pair. The data were analyzed following
the standard Whipple data analysis procedure \citep[see, e.g.][]{falcone04}. 
The log of the individual observations and 99.9~\% confidence upper limits, 
assuming an underlying Crab-like source spectrum with an underlying energy 
index of 2.49 \citep{hillas98}, are listed in Table \ref{Whipple_UL}. Combining 
all measurements results in a 99.9~\% confidence upper limit of $0.91 \times 
10^{-11}$~ergs~cm$^{-2}$~s$^{-1}$.

\begin{deluxetable}{cccc}
\tabletypesize{\scriptsize}
\tablecaption{Observation log and 99 \% confidence upper limits on 
the integrated $> 390$~GeV photon flux from the Whipple observations 
of 3C~66A, assuming a Crab-like spectrum. }
\tablewidth{0pt}
\tablehead{
\colhead{Date} & \colhead{Start time [UTC]} & 
\colhead{On-Source Live Time [min.]} & 
\colhead{99.9 \% UL [$10^{-11}$~ergs~cm$^{-2}$~s$^{-1}$]}
}
\startdata
030927	& 10:57	 &  27.73  &    2.05 \\
030928  & 09:04	 &  27.70  &    3.16 \\
030929	& 08:16	 &  27.75  &    2.73 \\
031002	& 08:58	 &  27.66  &    1.77 \\
031002	& 09:27	 &  27.73  &    2.35 \\
031002	& 10:25	 &  27.69  &    1.82 \\
031023	& 06:58	 &  27.55  &    6.19 \\
031024	& 07:07	 &  27.57  &    5.65 \\
031028	& 06:40	 &  27.55  &    6.35 \\
031028	& 07:08	 &  27.57  &    7.00 \\
031029	& 06:41	 &  27.57  &    6.62 \\
031029	& 08:15	 &  19.99  &    5.29 \\
031030	& 06:27	 &  27.61  &    4.39 \\
031101	& 07:17	 &  27.58  &    4.51 \\
031125	& 05:49	 &  27.55  &    6.32 \\
031128	& 04:42	 &  27.57  &    2.81 \\
031214	& 03:06	 &  27.62  &    4.87 \\
031217	& 03:00	 &  27.63  &    5.95 \\
031217	& 04:08	 &  27.54  &    5.23 \\
031217	& 04:37	 &  27.58  &    5.23 \\
031217	& 05:05	 &  27.55  &    6.71 \\
031217	& 05:33	 &  27.58  &    3.67 \\
031220	& 03:38	 &  27.63  &    2.60 \\
031220	& 05:09	 &  27.60  &    5.95 \\
031220	& 06:07	 &  27.66  &    4.45 \\
031223	& 02:48	 &  27.70  &    4.63 \\
031223	& 04:01	 &  27.71  &    4.03 \\
031223	& 04:31	 &  27.68  &    3.88 \\
031223	& 04:59	 &  27.70  &    2.94 \\
031225	& 04:35	 &  27.64  &    5.44 \\
040113	& 02:17	 &  27.58  &    5.23 \\
\enddata
\label{Whipple_UL}
\end{deluxetable}

\section{\label{variability}Spectral variability}

In this section, we will describe spectral variability
phenomena, i.e. the variability of spectral (and color) 
indices and their correlations with monochromatic source fluxes. 
As already mentioned in the previous section, no evidence for
spectral variability in the X-ray regime was found. Consequently,
we will concentrate here on the optical spectral variability as
indicated by a change of the optical color. In particular, our
observing strategy was optimized to obtain a good sampling of
the B -- R color index as a function of time. 

\begin{figure}
\plotone{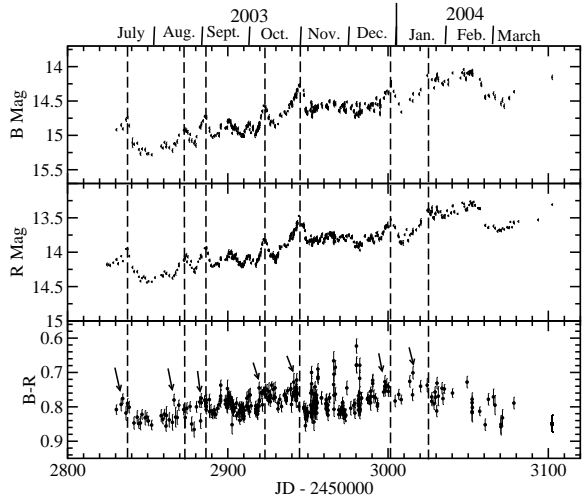}
\caption{Light curve of the B and R magnitudes and the B -- R color index 
of 3C~66A over the duration of the entire campaign. While the B and R
light curves are very well correlated, with no significant detection 
of a time delay, maxima of the spectral hardness (minima of B -- R) 
systematically precede B and R outbursts by a few days.}
\label{B_R_lc}
\end{figure}

\begin{figure}
\plotone{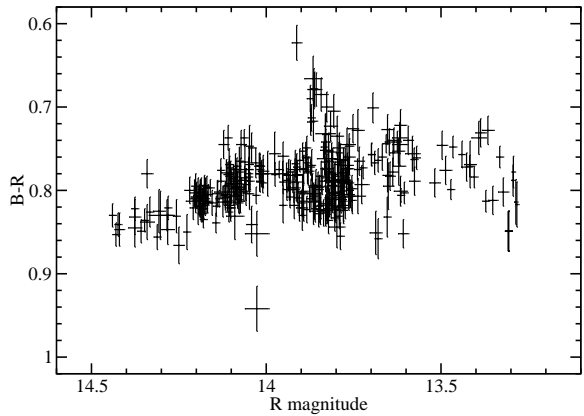}
\caption{Optical hardness -- intensity diagram for the complete data set
over the entire duration of the campaign. While a positive hardness-intensity
correlation seems to be present at low brightness, $R \gtrsim 14.0$, no
clear correlation can be found in brighter states. }
\label{hid_total}
\end{figure}

\begin{figure}
\plotone{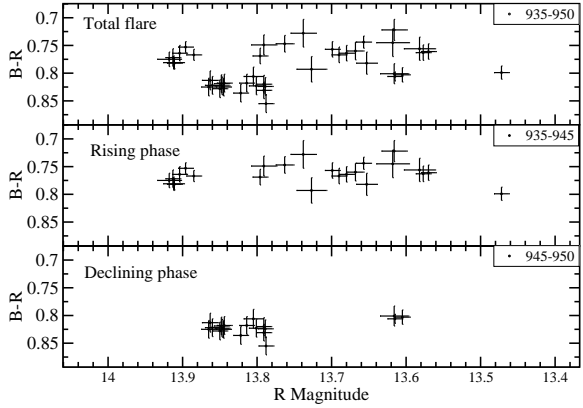}
\caption{Optical hardness -- intensity diagram for the flare around Nov. 1,
2003. The optical spectrum already begins to soften significantly while the 
R-band flux is still rising at $R \lesssim 13.6$. }
\label{hid_nov1}
\end{figure}

Based on our observation (see \S \ref{optical}) of no significant
variability on time scale less than $\sim 1$~hr, we extracted
B -- R color indices on measurements of B and R magnitudes separated
by less than 20 min. of each other. Fig. \ref{B_R_lc} shows a
comparison of the light curves of the B and R magnitudes and the 
B -- R color index. The figure reveals an interesting new result:
While the B and R light curves are well correlated, with no 
significant detection of a time lag from a discrete correlation 
function analysis (see \S \ref{crosscorrelations}), maxima of the
spectral hardness (i.e., minima of the value of B -- R) seem to
precede B and R band flux outbursts systematically by a few days.
In other words, the optical spectra harden at the onset of a major 
outburst, and continue to soften as the flare progresses through 
both the rising and the decaying phase. However, Fig. \ref{B_R_lc}
also indicates that the apparent lead time of the B -- R hardness 
maximum varies within a range of a few days. This might be the
reason that a cross-correlation analysis between the B -- R index
and the B or R light curves does not reveal a strong peak. In fact,
the DCF between those time series does never exceed values of $\sim
0.3$ at all.

Very interesting and intriguing is also the B -- R behaviour of 
the source around JD 2452955 -- JD 2452995, where no major flux
outbursts are observed, but several episodes of significant 
spectral hardening by $\Delta (B - R) \gtrsim 0.1$ (corresponding 
to spectral-index variations of $\Delta\alpha_{\rm opt} \gtrsim 0.2$) 
are observed.

In Fig. \ref{hid_total} we have plotted the B -- R index vs. R-band
magnitude for the entire data set. The plot shows that there is a
weak indication of a positive hardness-intensity correlation at low
flux states with $R \gtrsim 14.0$. At higher flux levels, no correlation
is apparent, which might be a consequence of the result found above,
that the B -- R hardness actually peaks during the rising phase of
individual outbursts. This is in contrast with recent results of
\cite{vagnetti03} who, on the basis of a smaller data set, have found
a consistent trend of B -- R hardening with increasing B-band flux,
independent of the actual flux value.

The optical spectral hysteresis visible in our data is further
illustrated in Fig. \ref{hid_nov1}, where we have plotted the 
B -- R vs. R hardness-intensity diagram for the example of one
individual outburst on Nov. 1, 2003 (MJD 52944). The top panel
shows the hardness-intensity diagram for the entire flare, which
is then split up into the rising and the decaying phase of the flare
in the middle and lower panels, respectively. The figure illustrates
that as the flare rises up to the peak R brightness, the spectrum
already begins to soften significantly as $R \lesssim 13.6$. Similar
trends are found for other optical flares as well. Possible physical
interpretations of this trend, along with detailed modelling of the
SED and spectral variability will be presented in a separate paper
(Joshi \& B\"ottcher 2005, in preparation).

\section{\label{crosscorrelations}Inter-band cross-correlations and time lags}

In this section, we investigate cross-correlations between the
measured light curves at different frequencies, within individual
frequency bands as well as broadband correlations between different
frequency bands. Because of relatively poor sampling of the radio
and IR light curves, our searches for correlations between different
radio and IR bands and between those wavelength bands with the optical
ones on short time scales compared to the duration of our campaign 
did not return conclusive results. Also, in agreement with our non-detection
of a periodicity on time scales of $\lesssim 50$~days, a discrete 
autocorrelation function \citep[DACF][]{ek88} analysis of individual 
optical light curves did not return results beyond artificial 
``periodicities''. Such artifacts can be attributed to the quasi-periodic, 
uneven time coverage which results in spurious detections of periodicities 
at multiples of days and multiples of $\sim 0.5$~d (the time delay between 
the peak coverage from the heavily contributing observatories in Europe 
and in East Asia). A discrete autocorrelation function analysis
of the R-band light curve also revealed secondary peaks at $\pm 4$~days
with a correlation coefficient close to 1.

The cleanest results of a DCF analysis are obviously expected for the
most densely sampled light curves, which we obtained in the B and R
bands. We performed DCFs between the R and B band light curves on a
variety of time scales with a variety of binning intervals, ranging
from 15 min. to 10 days. On intraday time scales, we find consistently
a sharp peak at 0 delay with all bin sizes we used, which indicates no 
evidence for a time delay between variability patterns in the R and B 
band reaching their peak fluxes in each band. The DCFs on short time 
scales are dominated by the artificial 0.5~d and 1~d periodicities 
mentioned above. The discrete correlation function between R and B band 
magnitudes on time scales of several days seems to indicate a strong correlation
at $\tau \sim -4$~days, corresponding to a lead of the B vs. the R band.
However, this might be a consequence of the probably artificial 4-day 
``periodicity'' of the R-band light curve discussed in the previous 
paragraph, which prevents us from making any claim about the detection
of a 4-day delay between the R and B band light curves on the basis of
our DCF analysis. 

\section{\label{spectra}Broad-band spectral energy distributions}

\begin{figure}
\plotone{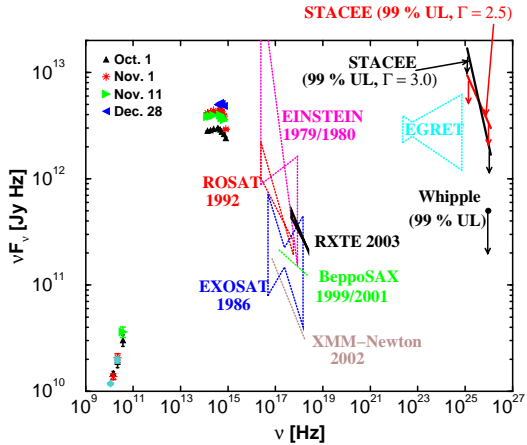}
\caption{Time-averaged spectral energy distributions of 3C~66A at various 
epochs during the core campaign period in 2003. The turquoise diamonds 
indicate the average of the RATAN-600 measurements from Oct. 11 and 14, 
2003. Historical soft X-ray spectra are plotted for comparison; the hard 
X-ray spectrum labelled ``RXTE 2003'' indicates the result of a power-law 
fit to the {\it RXTE}~PCA observations of this campaign. The historical 
(non-contemporaneous) average of the 5 EGRET pointings on the source is 
also included for reference. All data that are not contemporaneous with 
our campaign are indicated by dotted curves.}
\label{sed_plot}
\end{figure}

From the various flux measurements described in detail in the previous
sections, we can now compose contemporaneous spectral energy distributions 
(SEDs) of 3C~66A at various times during our campaign. Significant variability 
was only detected in the radio, IR, and optical bands. In those bands, we 
extracted SEDs for four epochs: During two major outbursts, around Nov. 1, 2003, 
and Dec. 28, 2003, a minor outburst around Oct. 1, 2003, as well as a rather 
quiescent state around Nov. 11, 2003. De-reddened optical and near-IR 
magnitudes were converted to fluxes using the zero-point normalizations 
of \cite{bessel98}. The resulting SEDs are plotted in Fig. \ref{sed_plot}. 
Due to the relatively poor sampling, radio and IR data were often not 
quite simultaneous with the optical spectra, which were extracted near 
the peaks of the invididual outbursts mentioned above. In this case, the
closest near-IR and radio data points were chosen. This led to time offsets
between the optical and the radio data points of up to $\sim 5$~d. Given the
relatively long time scale and moderate amplitude of variability at radio 
wavelengths, we are confident that this did not introduce serious distortions
of the low-frequency SED.

In addition to the data taken during our 2003/2004 campaign, we have included 
historical X-ray measurements, to indicate the degree of X-ray variability 
observed in this source and the historical average GeV $\gamma$-ray flux 
measured by the EGRET instrument on-board the {\it Compton Gamma-Ray Observatory} 
from 5 observations between Nov. 1991 and Sept. 1995 \citep{hartman99}. 

The shape of the time-averaged optical (U)BVRI spectra, together with the very
steep X-ray spectral index, indicates that the $\nu F_{\nu}$ peak of the 
synchrotron component of 3C~66A is typically located in the optical range.
The shape of our best-fit {\it RXTE} spectrum provides strong evidence that 
the synchrotron component extends far into the X-ray regime and intersects 
the high-energy component in the time-averaged SED of 3C~66A at $\gtrsim 10$~keV 
during our core campaign period. Due to the lack of simultaneous GeV $\gamma$-ray 
coverage and of a firm detection at $> 100$~GeV, we can not make a precise 
statement concerning the level of $\gamma$-ray emission during our campaign. 
However, if the historical EGRET flux is representative also for the time of 
our campaign, then the total energy  output in the low-frequency and the 
high-frequency components of the SED of 3C~66A are comparable, as is typical 
for the class of intermediate and low-frequency peaked BL~Lac objects.

\section{\label{parameter_estimates}Generic parameter estimates}

In this section, we discuss some general constraints on source parameters. 
We will first (\S \ref{sed_parameters}) focus on parameter estimates using 
the spectral energy distribution and optical intraday variability measurements,
most relevant to the innermost portion of the jet outflow, closest to the 
central engine. In \S \ref{VLBA_parameters}, we will use the results of our 
VLBA observations to estimate parameters of the relativistic flow on parsec 
scales.

\subsection{\label{sed_parameters}Parameters of the inner jet}

From the minimum variability time scale of $\Delta t_{\rm min} \sim 2$~hr, 
we can estimate the size of the emitting region as $R \lesssim c \, D \, 
\Delta t_{\rm min}$, where $D = \left( \Gamma \left[ 1 - \beta_{\Gamma} 
\cos\theta_{\rm obs} \right] \right)^{-1}$, where $\Gamma$ is the bulk Lorentz 
factor of the emitting region, $\beta_{\Gamma}$~c is the corresponding speed, 
and $\theta_{\rm obs}$ is the observing angle. This yields $R \lesssim 2.2 
\times 10^{15} \, \delta_1$~cm, where $\delta = 10 \, \delta_1 \sim 15$ as 
an estimate from the limits on superluminal motion and from previous modeling 
efforts, as mentioned in the introduction. 

An estimate of the co-moving magnetic field can be found by assuming 
that the dominant portion of the time-averaged synchrotron spectrum 
is emitted by a quasi-equilibrium power-law spectrum of electrons with 
$N_e (\gamma) = n_0 \, V_B \, \gamma^{-p}$ for $\gamma_1 \le \gamma 
\le \gamma_2$; here, $V_B$ is the co-moving blob volume. Based on the 
X-ray spectral index of $\alpha \approx 1.5$, we find a particle 
spectral index of $p \approx 4$. Since the synchrotron cooling time 
scale for X-ray emitting electrons might be much shorter than the 
dynamical time scale (see Eq. \ref{tau_sy}), the X-rays are likely 
to be produced by a cooled electron distribution. In this case, the 
index $p \approx 4$ corresponds to a distribution of electrons injected 
into the emission region with an original index $q \approx 3$. The 
normalization constant $n_0$ is related to the magnetic field through 
an equipartition parameter $e_B \equiv u_B / u_e$ (in the co-moving 
frame). Note that this equipartition parameter only refers to the 
energy density of the electrons, not accounting for a (possibly 
dominant) energy content of a hadronic matter component in the 
jet. Under these assumptions, the magnetic field can be estimated 
as described, e.g., in \cite{boettcher03}. Taking the $\nu F_{\nu}$ peak 
synchrotron flux $f_{\epsilon}^{\rm sy}$ at the dimensionless synchrotron 
peak energy $\epsilon_{\rm sy}$ as $\sim 5 \times 10^{-11}$~ergs~cm$^{-2}$~s$^{-1}$
at $\epsilon_{\rm sy} \approx 5 \times 10^{-6}$ and $R \approx 3.3 \times 
10^{15}$~cm, we find

\begin{equation}
B_{e_B} = 4.4 \, \delta_1^{-1} \, e_B^{2/7} \; {\rm G},
\label{B_eB}
\end{equation}
which yields $B_{e_B} \approx 2.9 \, e_B^{2/7}$~G for $\delta = 15$. 

We can further use this magnetic-field value to estimate the end points of 
the electron spectrum since the low-energy end, $\gamma_1$ might correspond
to the $\nu F_{\nu}$ peak of the synchrotron spectrum, and the synchrotron 
high-energy cutoff corresponds to $\gamma_2$. Generally, we find

\begin{equation}
\gamma \approx 3.1 \times 10^3 \, \nu_{15}^{1/2} \, \left( {\delta \over 15} \right)^{-1/2}
\, \left( {B \over 2.9 \, {\rm G}} \right)^{-1/2},
\label{gamma_estimate}
\end{equation}
where $\nu_{15}$ is the characteristic synchrotron frequency in units of 
$10^{15}$~Hz. With our standard parameters, this yields $\gamma_1 \approx 
3.1 \times 10^3$, and $\gamma_2 = 1.5 \times 10^5$, if the synchrotron cutoff 
occurs around 10~keV. We can also use this to estimate the synchrotron cooling
time scale of electrons in the observer's frame:

\begin{equation}
\tau_{\rm cool, sy}^{\rm obs} \approx 2.8 \times 10^3 \, \left( {\delta \over 15} \right)^{-1/2}
\, \left( {B \over 2.9 \, {\rm G}} \right)^{-3/2} \, \nu_{15}^{-1/2} \; {\rm s}.
\label{tau_sy}
\end{equation}
For optical frequencies, this yields observed cooling time scales of the order
of $\sim 2$~hr, in agreement with the observed minimum variability time scale. 
This, however, raises an important caveat: The observed minimum variability time
scale may, in fact, be a reflection of the electron cooling time scale rather than
the dynamical time scale, as we had assumed when choosing our estimate for the
source size $R$. For this reason, a more detailed future investigation of possible 
short-term variability at X-ray frequencies will be extremely important to resolve
this issue. If X-ray variability on shorter time scales than $\sim 2$~hr is found,
the emission region would then have to be more compact than the $R \sim 3 \times
10^{15}$~cm which we had estimated here, and X-ray spectral hysteresis patterns
could arise. In contrast, if the X-ray variability time scale is found to 
be consistent with the optical one, then it would have to be dominated by the 
dynamical time scale and thus be largely achromatic. Thus, in that case no 
significant X-ray spectral hysteresis would be expected. 

\subsection{\label{VLBA_parameters}Parameters of the pc-scale outflow}

In our analysis of the 22~GHz and 43~GHz VLBA maps of 3C~66A, we had found
a rather smooth jet, with only one of the 6 Gaussian components (C1) showing
evidence for superluminal motion of $\beta_{\rm app} = (8.5 \pm 5.6) \, 
h^{-1}$. If we consider the measured speed of the component C1 to be close 
to the maximum as observed under the superluminal angle given by 
$\cos(\theta_{\rm obs}^{\rm SL}) = \beta_{\Gamma}$, we can estimate 
the minimum Lorentz factor and maximum angle between the jet and our 
line of sight. With $h = 0.7$ and by taking the lower limit of the 
$\beta_{\rm app} \ge (8.5 - 5.6) \, h^{-1} = 2.9 \, h^{-1}$, we have 
a lower limit to the bulk Lorentz factor of $\Gamma \ge 4.3$ and an 
upper limit to the viewing angle of $\theta_{\rm obs} \le 27.2^o$.
The large error in $\beta_{\rm app}$ results in very weak constraints
on $\Gamma$ and $\theta_{\rm obs}$, and they should be considered
as very conservative estimates. Much more accurate values should be 
available after the analysis of all 9 epochs of the VLBA monitoring 
program (Savolainen et al. 2005, in preparation).

\begin{figure}
\plotone{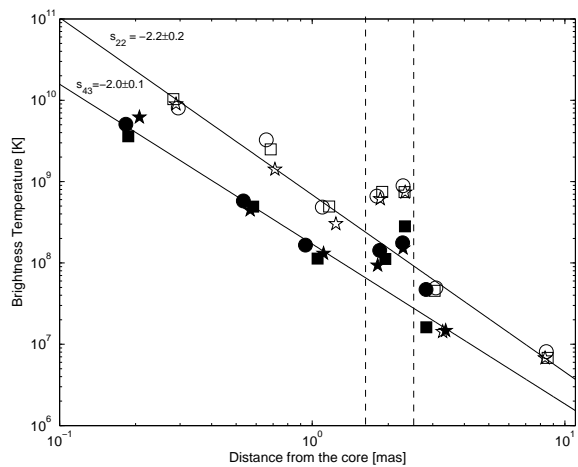}
\caption{VLBA brightness temperatures as a function of distance from the core, 
at 22~GHz (filled symbols) and 43~GHz (open symbols), for the three epochs: 
2003.78 (circles), 2003.83 (squares), 2004.08 (stars). The solid lines are 
our best-fit power-laws to the brightness-temperature profiles; the vertical 
dashed lines indicate the jet regions B2 and B3, where a significant deviation 
from the power-law profile is found, coincident with an apparent kink in the 
jet flow direction. }
\label{Tb_vs_rad}
\end{figure}

From the measured fluxes at 22 and 43~GHz, we have calculated the brightness 
temperatures along the jet. The largest brightness temperatures are found 
for the core, yielding lower limits of $T_b^{\rm core} \sim 10^{11}$~K (see 
Tab. \ref{core_Tb}). Here, we have to mention the caveat that the errors for 
the size of the core are very difficult to estimate, since our model-fitting 
yields very asymmetric probability density distributions for the core size. 
Essentially, the data is also consistent with a point source core, so we can 
only give an upper limit to the core size. Thus, no upper limit to the core 
brightness temperature is available. Our calculated best-fit core brightness
temperatures and corresponding lower limits are listed in Tab. \ref{core_Tb} 
and can be compared with an equipartion brightness temperature \citep{readhead94} 
for~3C 66A of $T_{\rm b, ep} = 6*10^{10}$~K. If the equipartition assumption 
holds, the minimum Doppler factor is around 2. The same result was obtained by
\cite{lv99} who used variability arguments to calculate the Doppler factor.
Because we do not have upper limits for the brightness temperature, we cannot 
calculate a more precise value for the Doppler factor. However, if the core 
brightness temperatures calculated from the best model-fit values are not 
too far from the truth, the core Doppler factor probably lies somewhere 
between 2 and 15 with the average best-fit value of 6.

We have also calculated the brightness temperatures for components other 
than the core at both 22 and 43~GHz. In Fig. \ref{Tb_vs_rad}, the component 
brightness temperatures are plotted against the component distance from the 
core. The figure illustrates that the points form two well-defined power 
laws with a different scaling at different frequencies. All three epochs 
are included in the plot (and coded by colors). The brightness temperature 
gradient along the jet is well described by a power law $T_b \propto r^s$ 
with $s = -2$. There is an exception to the power law between 1.8 and 2.3~mas
from the core, where the brightness temperature increases above the power law 
fit by a factor of 3 -- 5. This takes place in the jet region B, where the jet 
is clearly bending. This brightening could be a result of a temporary increase
of the Doppler factor caused by the bending of the jet towards our line of 
sight. An alternative interpretation could be given by impulsive particle 
acceleration in the region B by a standing shock wave caused by the interaction
of the continuous, relativistic outflow with the observed kink.

The power law gradient of the brightness temperature along the jet suggests
that 3C~66A exhibits a smooth pc-scale jet without any prominent particle
acceleration sites (shocks) other than the core and possibly the bend in
region B. If we assume that the magnetic field $B$, the electron density $N$ 
and the cross-sectional diameter $D$ of the jet can also be described by power 
laws: $N \propto r^n$, $B \propto r^b$ and $D \propto r^d$, the brightness 
temperature is expected to fall with $r$ as $T_b \propto r^s$ (with $s < 0$). 
If optically thin synchrotron emission and a constant Lorentz factor of the 
emitting electrons are assumed, $s = d + n + b (1 + \alpha)$, where $\alpha$ 
is the spectral index \citep[see, e.g.][]{kadler04}. If we know the values 
of $s$, $d$ and $\alpha$, we can calculate the relationship between $b$ and 
$n$, or, by assuming equipartition, the actual values of $b$ and $n$. We know 
now that $s = -2$, and the average spectral index is 
$\langle\alpha_{22 - 43}\rangle = 0.15$.

\begin{figure}
\plotone{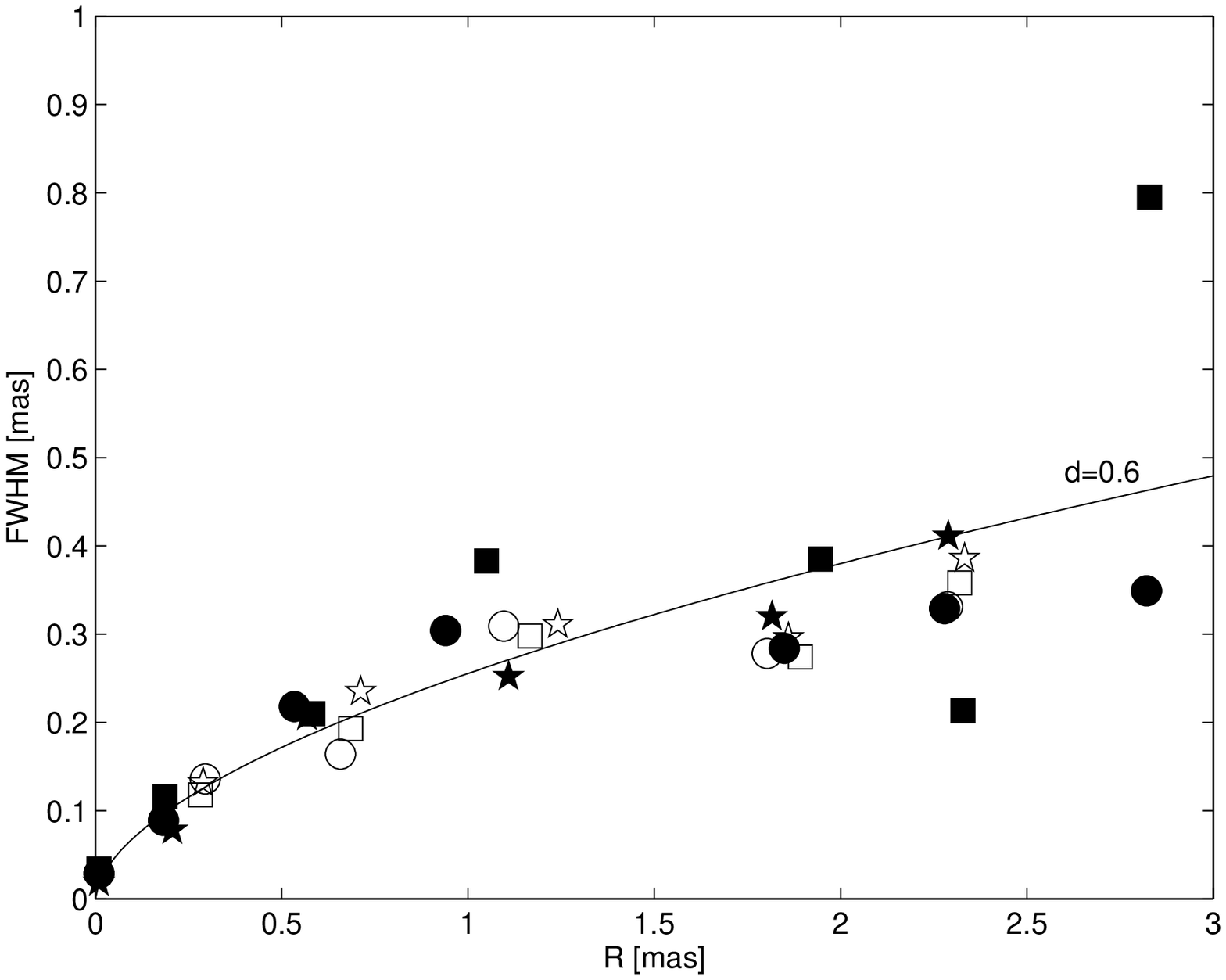}
\caption{Transverse diameter $D$ (FWHM of Gaussian component fits) of the
jet as a function of distance from the core for the inner 3~mas. The symbols
are the same as in Fig. \ref{Tb_vs_rad}. The solid curve illustrates the 
best-fit power law $D \propto r^d$, which yields $d = 0.6$.}
\label{opening_angle}
\end{figure}

We still need to find out the value of $d$. $d = 1$ would correspond to a 
constant opening angle (conical jet), but our imaging results allow us to 
estimate a more precise value from the data by plotting the component size 
versus distance from the core (see Fig. \ref{opening_angle}). 
Surprisingly, up to 2.5~mas from the core, the jet is very tightly 
collimated. A power-law fit to the observed correlation between the
lateral jet size and distance from the core yields $d = 0.6$. After 
2.5~mas, the jet opens up (or disrupts) and $d$ becomes larger than 1. 
This might be another indication that the angle between the jet and 
our line of sight is changing along the jet. The observed opening 
angle of the jet, $\psi_{\rm obs}$, is related to the intrinsic opening 
angle $\psi_{\rm int}$, through $\psi_{\rm obs} \approx \psi_{\rm int} / 
\sin(\theta_{\rm obs})$ for small values of $\psi_{\rm int}$ and 
$\theta_{\rm obs}$. Thus, the observed opening angle increases when 
the jet bends towards our line of sight, which is in accordance with 
the observed brightening of the jet in the region B. 

If equipartition holds, we may assume $n = 2 \, b$. Then, taking $d = 0.6$
yields $b = -0.8$ and $n = -1.7$. For the case of a conical jet ($d = 1$),
we find $n = -1.9$ and $b = -1.0$. If the equipartition assumption holds,
these results mean that in a conical jet, the magnetic field decays as
$B \propto D^{-1}$, and in a collimated jet ($d = 0.6$), it decays as
$B \propto D^{-1.3}$, implying a magnetic field that 
is predominantly transverse to the jet axis. This is in accordance with 
other observational results on the magnetic-field orientation in BL~Lac 
objects in general \citep[see, e.g.][]{gabuzda04}. A magnetic field 
being predominantly transverse to the jet agrees well with overall 
toroidal field configuration which is expected in Poynting flux dominated 
(PFD) jet models. Another possibility for creating predominantly transverse 
magnetic field is the enhancement of the transverse magnetic field component 
by a series of shocks, but this seems to be less likely in this case, since 
we do not observe bright knots in the jet, but a rather smooth flow.

\section{\label{summary}Summary}

We have observed 3C~66A in a massive multiwavelength
monitoring campaign from July 2003 to April 2004. 
Monitoring observations were carried out at radio,
infrared, optical, X-ray, and VHE $\gamma$-ray observations.
The main observational results of our campaign are:

\begin{itemize}

\item In the optical, several outbursts by $\Delta m
\sim 0.3$ -- $0.5$ over time scales of several days were
observed.

\item Optical intraday microvariability ($\Delta m \lesssim 
0.05$) on time scales of $\Delta t_{\rm var} \sim 2$~hr was
detected. 

\item No clear evidence for periodicity was found, but a
quasi-regular sequence of several major outbursts separated 
by $\sim 50$ - 57~d was identified. 

\item Large optical flares (on time scales of several days)
seem to exhibit optical spectral hysteresis, with 
the B - R
hardness peaking several days prior to the 
R- and B-band flux peaks.

\item The 3 -- 10~keV X-ray spectrum is best fitted with a single
power-law with energy index $\alpha = 1.47 \pm 0.56$, indicating 
that the transition
between the synchrotron and the high-energy 
components occurs
at photon energies of $\gtrsim 10$~keV.

\item Radio VLBA monitoring observations reveal a rather
smooth jet with no clearly discernible knots or hot spots
except for moderate brightening in the region B where the
jet is bending.

\item Decomposition of the VLBA radio structure into Gaussian
components reveals superluminal motion only for one out of 6
components. Its apparent speed is $\beta_{\rm app} = (8.5 \pm
5.6) \, h^{-1} = (12.1 \pm 8.0)$ for $h = 0.7$.

\item The radio brightness temperature profile along the jet,
along with its observed geometry, $D \propto r^{0.6}$, suggest
a magnetic-field decay $B \propto D^{-1.3}$, indicating a
predominantly perpendicular magnetic-field orientation.

\item STACEE observations revealed a $2.2 \, \sigma$ excess,
which provided strict upper limits at $E_{\gamma} \gtrsim 150$~GeV. 
Additional VHE $\gamma$-ray limits at $E_{\gamma} > 390$~GeV 
resulted from simultaneous Whipple observations.

\item The broadband SED of 3C~66A during our campaign suggests
that the synchrotron component peaks in the optical and extends
far into the X-ray regime, out to at least $\sim 10$~keV. 

\item The following parameters of the synchrotron emission region 
near the core can be estimated:
\begin{eqnarray}
\delta &\approx& 15 \hss \cr
R &\approx& 3.3 \times 10^{15} \; {\rm cm} \hss \cr
B &\approx& 2.9 \, \epsilon_B^{2/7} \; {\rm G} \hss \cr
\gamma_1 &\approx& 3.1 \times 10^3 \hss \cr
\gamma_2 &\approx& 1.5 \times 10^5 \hss \cr
p &\approx& 4 \hss 
\label{parameter_summary}
\end{eqnarray}

\item No X-ray variability was detectable by {\it RXTE} on a $\sim 1$~day
time scale. Future X-ray observations with more sensitive X-ray detectors 
will be important to probe for rapid X-ray variability and X-ray spectral
hysteresis in order to put more stringent constraints on the source size,
the nature of the variability mechanism, and the composition and energetics
of the emitting plasma in the jet.

\end{itemize}

\acknowledgments
The work of M. B\"ottcher and M. Joshi at Ohio University and of G. Fossati and
I. A. Smith at Rice University was partially supported through NASA's RXTE guest 
observer program, award no. NNG~04GB13G.
The European Institutes belonging to the ENIGMA collaboration acknowledge EC 
funding under contract HPRN-CT-2002-00321.
The UMRAO is partially supported by funds from the NSF and from the University
of Michigan's Department of Astronomy.
The work of T. Savolainen was partly supported by the Finnish Cultural Foundation.
The St.-Petersburg group was supported by Federal Programs ``Astronomy"
(grant 40.022.1.1.1001) and ``Integration" (grant B0029).
J. H. Fan's work is partially supported by the National 973 project 
(NKBRSF G19990754) of China, the National Science Fund for Distinguished 
Young Scholars (10125313).
V. Hagen-Thorn and V. Larionov acknowledge support from the Russian Foundation 
for Basic Research, project 05-02-17562.
The work at the Mt. Abu Infrared Observatory was supported by the DOS,
Govt. of India.
W. Cui nd D. Able gratefully acknowledge support from the Department of Energy.
The RATAN-600 observations were partly supported by the NASA JURRISS Program
(grant W-19611) and the Russian Foundation for Basic Research (grant 05-02-17377).

\end{document}